\documentclass[prb,twocolumn]{revtex4-2}
\usepackage{amsmath,amssymb,bm}
\usepackage{hyperref}
\usepackage[]{graphicx}
\usepackage{epstopdf}
\usepackage{latexsym}
\usepackage{subfigure}
\usepackage[usenames, dvipsnames]{color}
\usepackage[usenames, dvipsnames]{xcolor}
\usepackage{natbib}
\usepackage{braket}
\usepackage{float}
\usepackage[normalem]{ulem}
\usepackage{comment}
\usepackage{mathtools}
\usepackage{array}
\usepackage{tabu}
\usepackage{multirow}
\usepackage{chemformula}
\usepackage{svg}

\begin{document}
\title{Competition between Neel, Haldane nematic, plaquette valence bond solid, and $(\pi,\pi)$ valence bond solid phases in SU(N) analogs of $S=1$ square-lattice antiferromagnets
}
\author{Souvik Kundu}
\author{Nisheeta Desai}
\author{Kedar Damle}
\affiliation{Department of Theoretical Physics, Tata Institute of Fundamental Research,
Mumbai, India}
\begin{abstract}
We use stochastic series expansion (SSE) quantum Monte Carlo (QMC) methods to study the phases and transitions displayed by a class of sign-free designer Hamiltonians for SU($N$) analogs of spin $S=1$ quantum antiferromagnets on the square lattice. The SU($N$) spins are generators of the single-row two-column representation (complex conjugate of single-row two-column representation) on $A$ ($B$) sublattices of the square lattice, and the Hamiltonian is designed to explore the competition between the nearest neighbour antiferromagnetic exchange couplings $J$ and four-spin interactions $Q$ that favor a plaquette-ordered valence bond solid (p-VBS) ground state. We find that this state is indeed established at large $Q/J$ for all $N >  3$. For $3< N \leq 9$, the ground state exhibits a direct first order quantum phase transition from a small-$Q/J$ N\'eel ordered antiferromagnetic state to this large-$Q/J$ p-VBS state. The ground state at $Q/J=0$ for $N \geq 10$ has been previously reported to be a valence bond nematic state, dubbed the Haldane nematic in recent literature. For small nonzero $Q/J$ and $N \geq 10$, we additionally find an unusual intermediate state in which the bond energy has a Bragg peak at wavevector $(\pi,\pi)$ with no accompanying Bragg peaks at wavevectors $(\pi,0)$ and $(0, \pi)$. This $(\pi, \pi)$ state also appears to be metastable at $Q/J =0$, as evidenced by low temperature histograms of the Haldane nematic and $(\pi, \pi)$ order parameters.
Deep in the p-VBS phase of the ground state phase diagram, we find the temperature-driven melting of the p-VBS order is in the Ashkin-Teller universality class. In this regime, we identify an interesting signature of Ashkin-Teller criticality in bond correlations at wavevector $(\pi, \pi)$; this is in addition to the expected critical fluctuations of the conventional p-VBS order parameter at wavevectors $(\pi,0)$ and $(0,\pi)$.

\end{abstract}
\maketitle

\section{Introduction}

Quantum antiferromagnetism provides a rich arena for the study of several phenomena of interest in condensed matter physics~\cite{auerbach1998,Sachdev2008,wen2004,moessner2021}. For instance, the Haldane-gapped ground state of spin $S=1$ antiferromagnetic chains~\cite{Haldane_prl1983,Affleck_prl1985,Affleck_etal_prl1987,Affleck_etal_cmp1988} represents the prototypical example of topological order, with characteristic ground state degeneracy that is robust to perturbations in the microscopic Hamiltonian. 

More generally, the competition between magnetic order (favoured by the exchange couplings) and magnetically disordered ground states (favoured by quantum fluctuations) can lead to a variety of interesting phases and phase transitions. In some cases, it can give rise to spin liquid phases with topological order.~\cite{Alet_2006,wen2004} In addition, geometric frustration effects can lead to even richer physics~\cite{Savary_Balents_2017}. The large number of experimentally-studied Mott insulating compounds that realize some of these phenomena is another reason for the theoretical interest in these systems.~\cite{Balents_nature2010,Broholm_etal_Science2020,Knolle_Moessner_review2019} 

A major theoretical challenge is the paucity of reliable computational methods for characterizing the ground state or low-temperature behavior of all but the simplest of the experimentally-relevant Hamiltonians that display such interesting phenomena; in large part, this is due to the sign-problem faced by quantum Monte Carlo (QMC) simulations when the interactions are frustrated. One way around this is the study of so-called ``designer Hamiltonians'' constructed to be both computationally tractable and physically relevant (in the sense that their low-temperature phase diagrams display the phases and transitions of interest)~\cite{Sandvik_prl2007,lou2009:q3model,sandvik2010:deconf,melko2008:fan, Sandvik_prb2012, Pujari_Damle_Alet_prl2013, Pujari_Alet_Damle_prb2015, Kaul_prb2012, Kaul_Sandvik_prl2012, Block_Melko_Kaul_prl2013, Harada_etal_prb2013,Jiang_JStatMech2008,Kuklov_2006,Kun_prl2013,Kaul_prb2011,Shao2016:science,Takahashi_Sandvik_prr2020,Kaul_Melko_Sandvik_2013}.

This typically involves replacing the physical ring-exchange couplings (expected to be relevant for the physics of spin $S=1/2$ Mott insulators with a not-very-large charge gap) by computationally tractable multi-spin interactions and/or enlarging the symmetries of the problem from SU($2$) to SU($N$) or other symmetry groups~\cite{Sandvik_prl2007,Affleck_prl1985,auerbach_arovas_prb1988,Read_Sachdev_NucPhysB,Read_Sachdev_prb1990,Read_Sachdev_prl1989}. QMC algorithms have also been developed for the detailed study of higher spin $S \geq 1$ systems~\cite{Kawashima_GubernatisPRL1994,Todo_Kato_prl2001} as well as their SU($N$) generalizations~\cite{Kawashima_Tanabe_prl2007,Okubo_etal_prb2015}  using the so-called ``mini-spin'' representations~\cite{Affleck_etal_cmp1988,Affleck_etal_prl1987}. Interestingly, the remarkable degree of control achieved in modern cold-atom experiments raises the possibility of experimental ``emulation'' of some of these designer systems, adding to the interest in their physics~\cite{Taie_2022, wu_2006, Cazalilla_etal_2009, Gorshkov_natphys2010, Cazalilla_2014, Honerkamp_prl2004, Hermele_prl2009, DelRe_pra2018, Tusi_2022}. 

This approach has been quite fruitful in the past: For instance, the transition from the N\'eel phase to a columnar valence bond solid (cVBS) phase has been studied extensively for SU($2$) symmetric $S=1/2$ models as well as their generalizations to SU($N$)~\cite{Sandvik_prl2007,lou2009:q3model,sandvik2010:deconf,Kaul_prb2015,melko2008:fan, Sandvik_prb2012, Pujari_Damle_Alet_prl2013, Pujari_Alet_Damle_prb2015, Kaul_prb2012, Kaul_Sandvik_prl2012, Block_Melko_Kaul_prl2013, Harada_etal_prb2013,Jiang_JStatMech2008,Kuklov_2006,Kun_prl2013,Kaul_prb2011,Shao2016:science} motivated by the proposal of deconfined criticality~\cite{senthil2004:science}. Additionally, it has been used to study various kinds of impurity effects in valence bond ordered and magnetically ordered phases, as well as at the phase transitions between such phases~\cite{Banerjee_Damle_Alet_prb2010, Banerjee_Damle_Alet_prb2011, Sanyal_etal_prb2011, Sanyal_etal_prb2012, Liu_etal_prx2018, Liu_etal_prb2020, Shu_etal_prb2016, Shu_etal_prb2018}. \begin{figure}[t]
	\centering
	\includegraphics[width=\linewidth]{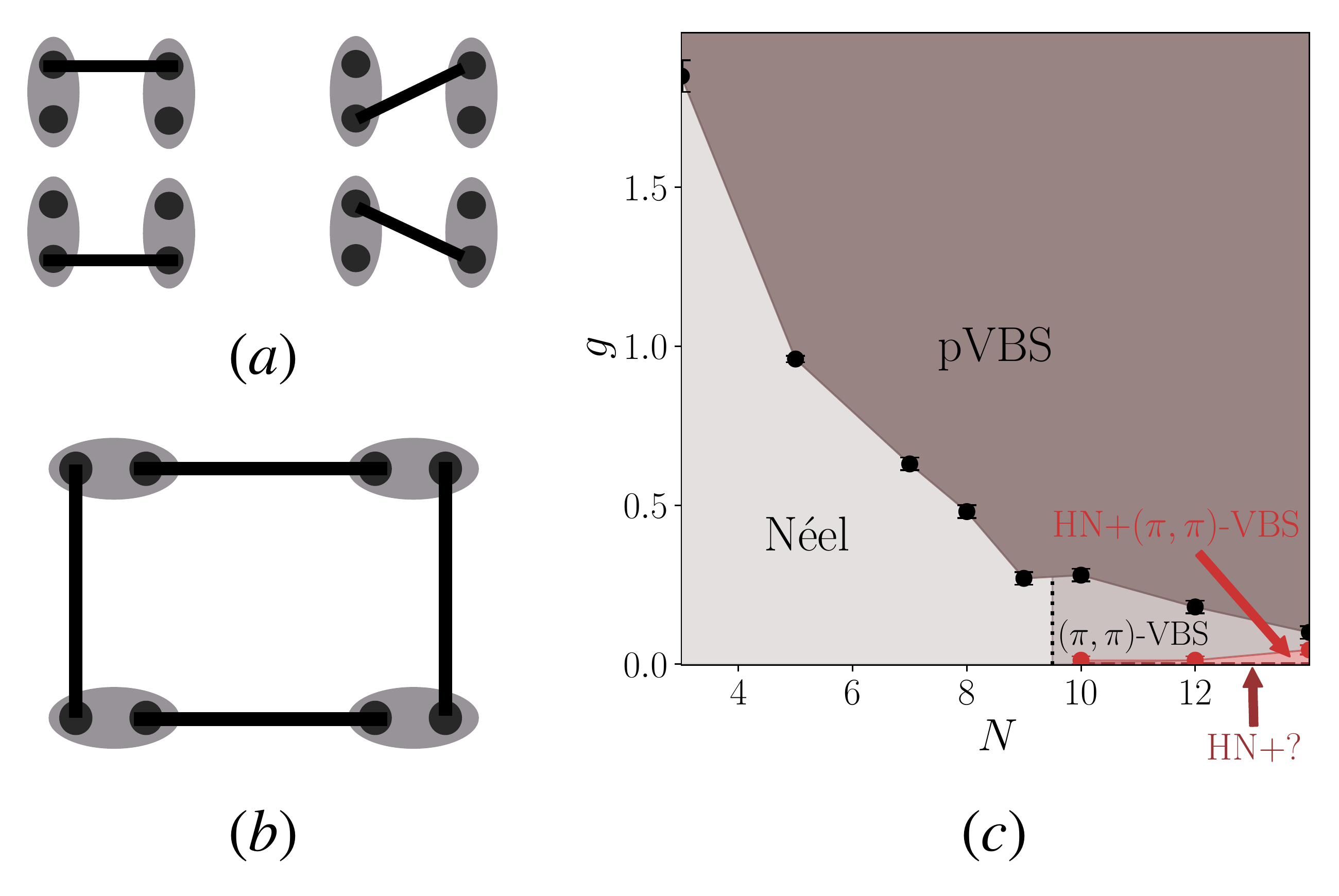}
	\caption{\label{fig:phase_diagram} 
 A lattice site is indicated by the gray bubble which contains $2$ symmetrized minispins . Minispins on $A$ ($B$) sublattice sites are generators of the single-row two-column representation (complex conjugate of the single-row two-column representation). Solid lines connecting two minispins denote SU($N$) singlet projectors that project to the SU($N$) singlet state formed by a pair of minispins on opposite sublattices.  Minispin picture for (a) the Heisenberg interaction, explicitly showing all four components generated by the action of the symmetrizers at each physical site (b) the plaquette interaction, showing only one of the sixteen different terms generated by the action of the symmetrizers (see Sec.~\ref{sec:Model} for a detailed account of the minispin formalism and definition of the symmetrizer). (c) Phase diagram as a function of $N$ and $g = Q/J$, where $Q$ is the strength of the plaquette interaction and $J$, the strength of the Heisenberg interaction. For $N\leq9$,  the $g=0$ groundstate is the N\'eel state, while the large $g$ state is a plaquette valence bond solid (p-VBS) state. The transition from N\'eel order to pVBS order on increasing $g$ is of the first order type. See Sec.~\ref{sec:ResultsSmallerN} for details. For $N \geq 10$, the $g=0$ ground state was previously identified to be a Haldane nematic (HN) state~\cite{Okubo_etal_prb2015}.  For these large values of $N$  we find intriguing phase coexistence phenomena at small $g$ close to zero, indicating the presence of a metastable state with $(\pi, \pi)$ VBS order. For intermediate values of $g$, we find that this $(\pi, \pi)$ VBS state wins over the Haldane nematic state, while for large values of $g$, the system has p-VBS order.  See Sec.~\ref{sec:ResultsLargerN} for details.}
\end{figure}

Using the minispin representation mentioned earlier, a state with bond nematic order predicted previously~\cite{Read_Sachdev_NucPhysB,Read_Sachdev_prb1990,Read_Sachdev_prl1989} was also identified in QMC studies of SU($N$) analogs of the spin $S=1$ square lattice antiferromagnet~\cite{Okubo_etal_prb2015}. More recently, several interesting SU($2$) symmetric designer Hamiltonians with $S \geq 1$ have also been introduced, and used in the $S=1$ case to understand the competition between the N\'eel phase, the columnar valence bond solid (cVBS) state, and a state with this kind of bond nematic order (dubbed the Haldane nematic state in this recent literature)~\cite{Desai_Kaul_prl2019,Wildeboer_etal_prb2020} on the square lattice.

Here, we use stochastic series expansion (SSE) quantum Monte Carlo (QMC) methods to study the phases and transitions displayed by a class of  sign-free designer Hamiltonians for SU($N$) analogs of spin $S=1$ quantum antiferromagnets on the square lattice. The SU($N$) spins are generators of the single-row two-column representation (complex conjugate of single-row two-column representation) on $A$ ($B$) sublattices of the square lattice~\cite{Read_Sachdev_NucPhysB,Read_Sachdev_prb1990,Read_Sachdev_prl1989,Okubo_etal_prb2015}, and the Hamiltonian is designed to explore the competition between the nearest neighbour antiferromagnetic exchange couplings $J$ and four-spin interactions $Q$ that favor a plaquette-ordered valence bond solid (p-VBS) ground state. 	\begin{figure}[t]
		\centering
		\subfigure[]{\includegraphics[width=0.47\linewidth]{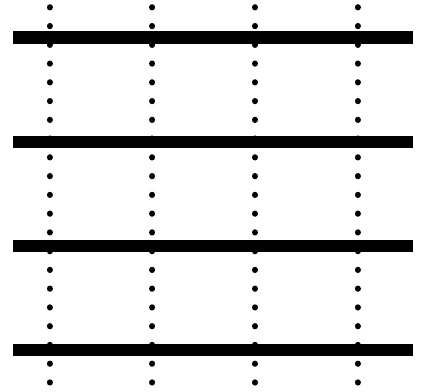}}
		\subfigure[]{\includegraphics[width=0.47\linewidth]{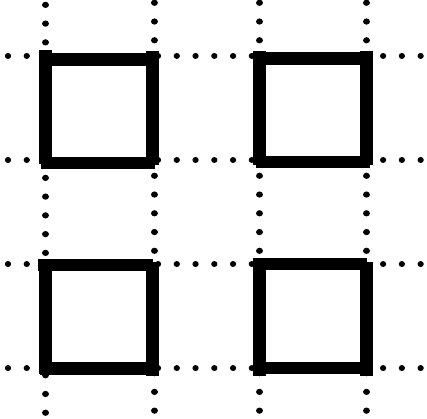}}
		\caption{\label{fig:quantum_phases} 
	(a) Schematic of the two-fold symmetry breaking Haldane nematic (HN) phase with bond nematic order. This breaks the rotational symmetry of the square lattice, leaving all translational and spin symmetries intact. (b) Four-fold symmetry breaking plaquette valence bond solid (p-VBS) phase, which breaks the translational symmetry of the square lattice, keeping intact spin symmetries and the symmetry of $90^{\text{o}}$ lattice rotations about a plaquette center. The dark bonds schematically represent higher values for the expectation value of the singlet projection operator that projects to the SU($N$) singlet state formed from two neighbouring spins on opposite sublattices. Note that the p-VBS state leads to Bragg peaks at wavevector $(\pi, 0)$ in the horizontal bond energy and wavevector $(0,\pi)$ in the vertical bond energy. It also leads to a subdominant Bragg peak at wavevector $(\pi, \pi)$ in the plaquette energy.}
	\end{figure}

We find that this p-VBS state is indeed established at large $Q/J$ for all $N >  3$. For $3< N \leq 9$, the ground state exhibits a direct first order quantum phase transition from a small-$Q/J$ N\'eel ordered antiferromagnetic state to this large-$Q/J$ p-VBS state. The ground state at $Q/J=0$ for $N \geq 10$ has been previously reported~\cite{Okubo_etal_prb2015} to be the Haldane nematic state described above. For small nonzero $Q/J$ and $N \geq 10$, we additionally find an unusual intermediate phase in which the plaquette energy has a Bragg peak at wavevector $(\pi,\pi)$, {\em with no accompanying Bragg peaks in the horizontal and vertical bond energies} at wavevectors $(\pi,0)$ and $(0, \pi)$ respectively.

This is strikingly different from the p-VBS phase at higher $Q/J$, in which a {\em subdominant} $(\pi, \pi)$ Bragg peak in the plaquette energies arises {\em as a consequence of the simultaneous ordering of the horizontal and vertical bond energies} at  wavevectors $(\pi,0)$ and $(0,\pi)$ respectively. Furthermore, a detailed study of the histograms of Haldane nematic, plaquette and $(\pi,\pi)$ order parameters suggests that this $(\pi, \pi)$ ordered state is also metastable in the $Q/J \rightarrow 0$ limit. A schematic of the exchange coupling and four-spin interaction terms of this Hamiltonian is shown in Fig.~\ref{fig:phase_diagram}, while Fig.~\ref{fig:quantum_phases} provides a schematic representation of the p-VBS and Haldane nematic phases.
	\begin{figure}[t]
		\centering
		\includegraphics[width=\linewidth]{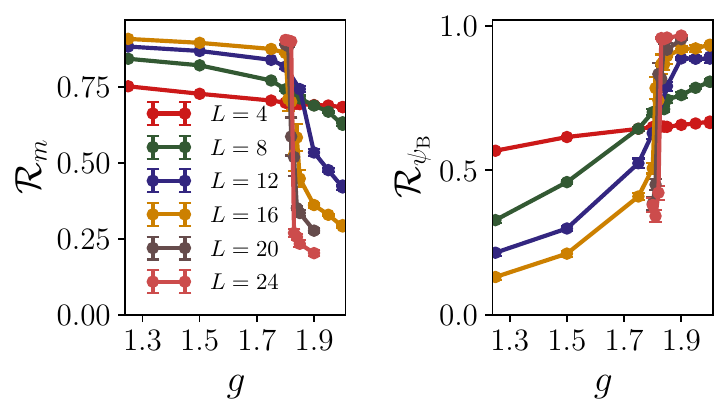}
		\caption{\label{fig:corrratios_N3} Dimensionless ratios constructed from N\'eel and pVBS order parameters, $\mathcal{R}_{m^2}$ and $\mathcal{R}_{\psi_{\rm B}^2}$, (see Sec.~\ref{sec:Methodsandobservables} for details of the definition) plotted as a function of $g$ for different system sizes ($L$) at $N=3$. In both cases, curves for different $L$ cross at $g\approx1.83$ indicates a direct transition between the two phases. See Sec.~\ref{sec:ResultsSmallerN} for details.}
	\end{figure}

Deep in the p-VBS phase of the ground state phase diagram, we find the temperature-driven melting of the p-VBS order is in the Ashkin-Teller universality class. In this regime, we identify an unusual signature of Ashkin-Teller criticality in bond correlations at wavevector $(\pi, \pi)$; this is in addition to the expected critical fluctuations of the conventional p-VBS order parameter at wavevectors $(\pi,0)$ and $(0,\pi)$. Interestingly, this is specific to the plaquette ordered case and very different from the better-understood subdominant critical fluctuations of the nematic order parameter at an Ashkin-Teller transition to a low temperature phase with {\em columnar} VBS order.~\cite{Ramola_Damle_Dhar_prl2015}. The resulting picture of the complete phase diagram is summarized in Fig.~\ref{fig:phase_diagram}.

The rest of the paper is organized as follows: In Sec.~\ref{sec:Model}, we discuss the model Hamiltonians studied, and explain how it may be viewed as one possible class of SU(N) generalizations of $S=1$ quantum antiferromagnet. In Sec.~\ref{sec:MethodsandObservables}, we provide an account of our QMC methods and the observables we study. In Sec.~\ref{sec:ResultsSmallerN}, we  describe our results for smaller values of $N$,    {\em i.e.} $N \leq 9$. In Sec.~\ref{sec:ResultsLargerN}, we describe our results for larger values of $N$, {\em i.e.} $N > 9$. Sec.~\ref{sec:AshkinTeller} focuses on our study of the temperature driven melting of p-VBS order deeper in the p-VBS ordered phase for both large and small values of $N$. We conclude with a brief discussion in Sec.~\ref{sec:Discussion}

\section{Model}
\label{sec:Model}

Following earlier work~\cite{Affleck_prl1985,auerbach_arovas_prb1988,Read_Sachdev_NucPhysB,Read_Sachdev_prb1990,Read_Sachdev_prl1989, Harada_Kawashima_Troyer_prl2003, Beach_etal_prb2009,Kawashima_Tanabe_prl2007,Okubo_etal_prb2015}  
we consider a generalization of the SU($2$) symmetric square lattice Heisenberg antiferromagnet, in which the spins are replaced by SU($N$) generators. Specifically, our spins on the $A$ sublattice of the square lattice are generators of the single-row two-column representation, while those on the $B$ sublattice are generators in the complex conjugate of this representation. These degrees of freedom can be thought of as being the SU($N$) analogs of spin $S=1$ moments. The validity of this interpretation is particularly clear when thought of in the language of our QMC calculations. These calculations use the SU($N$) generalization~\cite{Kawashima_Tanabe_prl2007,Okubo_etal_prb2015} of a computational scheme~\cite{Kawashima_GubernatisPRL1994,Todo_Kato_prl2001,Desai_Kaul_prl2019} that works within the minispin formalism developed earlier for understanding Haldane gapped spin $S=1$ antiferromagnets in one dimension~\cite{Affleck_etal_cmp1988,Affleck_etal_prl1987}. In this formalism, one views a spin $S \geq 1$ moment as being formed by symmetrizing $2S$ spin $S=1/2$ moments (these are the ``minispins''), {\em i.e.} keeping only the highest spin multiplet from the angular moment addition of $2S$ spin $S=1/2$ moments. Since the single-row two-column representation of SU($N$) is obtained by keeping only the symmetric part of the tensor product of two fundamental representations, we see immediately that the same minispin formalism carries over unchanged to the SU($N$) case. This justifies viewing our degrees of freedom as SU($N$) analogs of $S=1$ moments. For a pictorial depiction of the minispin representation, see Fig.~\ref{fig:phase_diagram}
\begin{figure}[t]
\centering
\subfigure{\includegraphics[width=\linewidth]{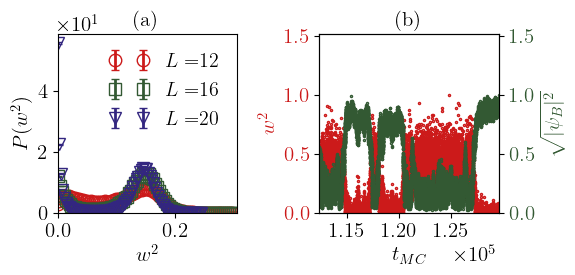}}
\subfigure{\includegraphics[width=\linewidth]{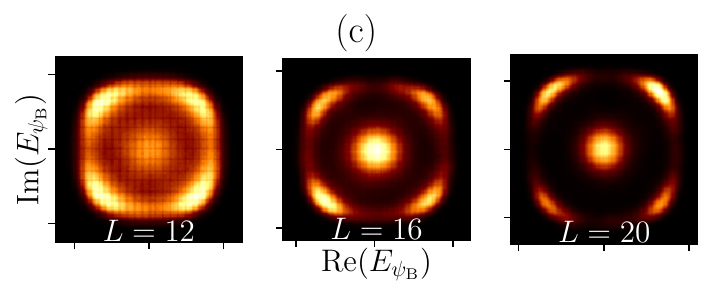}}   
\caption{\label{fig:hist_transition_N3} Order parameter histograms for $N=3$ close to the N\'eel-pVBS transition at $g \approx 1.83$ and $\beta=12$ (zero temperature limit) (a) Histograms of spin stiffness, $\rho_s$, showing two peaks getting sharper with increasing system size at $g=1.83,1.81,1.81$ for $L=12,16,20$ respectively.  (b) Monte-Carlo time series of $w^2$ and $\sqrt(|\psi_{\rm B}|^2)$ for $L=20$ with the coupling parameter $g=1.82$. The values of each estimators are normalized such that the maximum value is unity. It shows switching between N\'eel and VBS phase throughout the MC sampling history near the transition point. (c) Joint histograms of the real and imaginary part of the estimator of the complex order parameter $\psi_{\rm B}$, shows five peaks near the transition (at $g=1.87,1.83,1.82$ for $L=12,16,20$ respectively): one at the origin and four at finite values of $Re(E_{\psi_\mathrm{B}})$ and $Im(E_{\psi_\mathrm{B}})$ corresponding to pVBS order. The nature of the histograms show co-existence of pVBS and N\'eel order indicating a first order transition between the two phases. See Sec.~\ref{sec:ResultsSmallerN} for details.}
\end{figure}
\begin{figure}[t]
	\centering
	\includegraphics[width=\linewidth]{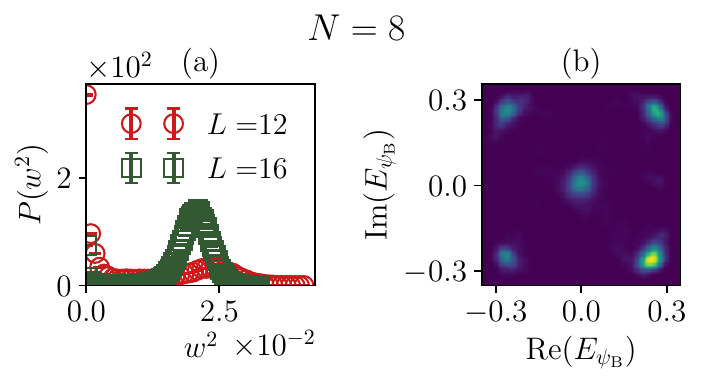}
	\includegraphics[width=\linewidth]{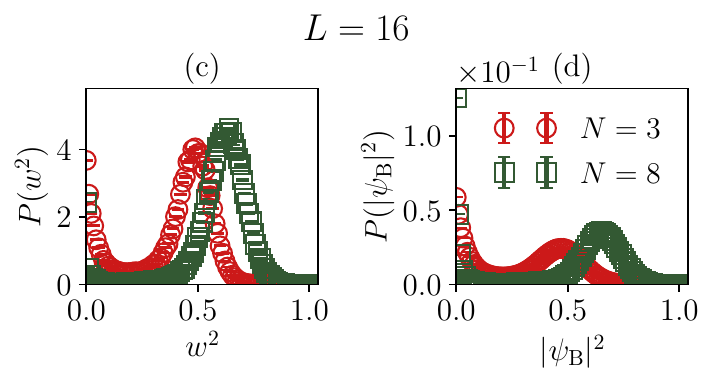}
	\caption{\label{fig:hist_transition_N8}(a) Histograms of the spin stiffness estimator and (b) joint histogram of the estimator of  VBS order parameter $\hat{\psi_{\rm B}}$ in the complex plane $Re(E_{\psi_\mathrm{B}})\mbox{-} Im(E_{\psi_\mathrm{B}})$,  close to the N\'eel-pVBS transition at $g\approx0.47$ for $N=8$ indicating strong first order nature of this transition at $\beta=48$ (zero temperature limit). The comparison of histograms of (c) $w^2$ and (d) $|\psi_{\rm B}|^2$ at the transition for the same system size, $L=16$, for $N=3$ and $N=8$ show clearly that the double-peaked structures are more pronounced for $N=8$. See Sec.~\ref{sec:ResultsSmallerN} for details}
\end{figure}

The antiferromagnetic Heisenberg interaction between two $S=1/2$ moments $S_1$ and $S_2$ can be written (up to an additive constant) as
\begin{displaymath}
-J P_{12}
\end{displaymath}
where $P_{12}$ is the projection operator that projects to the SU($2$) singlet sector of this two spin system.
With this as the template, the analogous SU($N$) symmetric interaction terms can be constructed using projectors that project to the SU($N$) singlet subspace in the tensor product of the local Hilbert spaces of one minispin on a $A$ sublattice site (carrying the fundamental representation) and another on a $B$ sublattice site (carrying the complex conjugate of the fundamental representation). We write the normalized SU($N$) singlet state formed from two such minispins $ia$ (living on an $A$ sublattice site) and $jb$ (living on a $B$ sublattice site) as
\begin{equation}
 |S_{ij}^{a,b}\rangle = \frac{1}{\sqrt N} \sum_{\alpha_{ia} = \bar{\alpha}_{jb}} |\alpha_{ia} \bar{\alpha}_{jb}\rangle \; ,
 \label{eq:suNsinglet}
\end{equation}
where the index $\alpha_{ia}$ carries the fundamental representation, while $\bar{\alpha}_{jb}$ carries the complex conjugate of the fundamental representation, and the sum is from $1$ to $N$. Using this to define the projector
\begin{eqnarray}
{\mathcal S}_{ij}^{a,b} &=& |S_{ij}^{a,b}\rangle \langle S_{ij}^{a,b}| 
\end{eqnarray}
that projects to the singlet formed by these two minispins, we may write the antiferromagnetic Heisenberg interaction between these two SU($N$) minispins as
\begin{eqnarray}
-{\mathcal S}_{ij}^{a,b} \; .
\end{eqnarray}
\begin{figure}
	\centering
	\includegraphics[width=\linewidth]{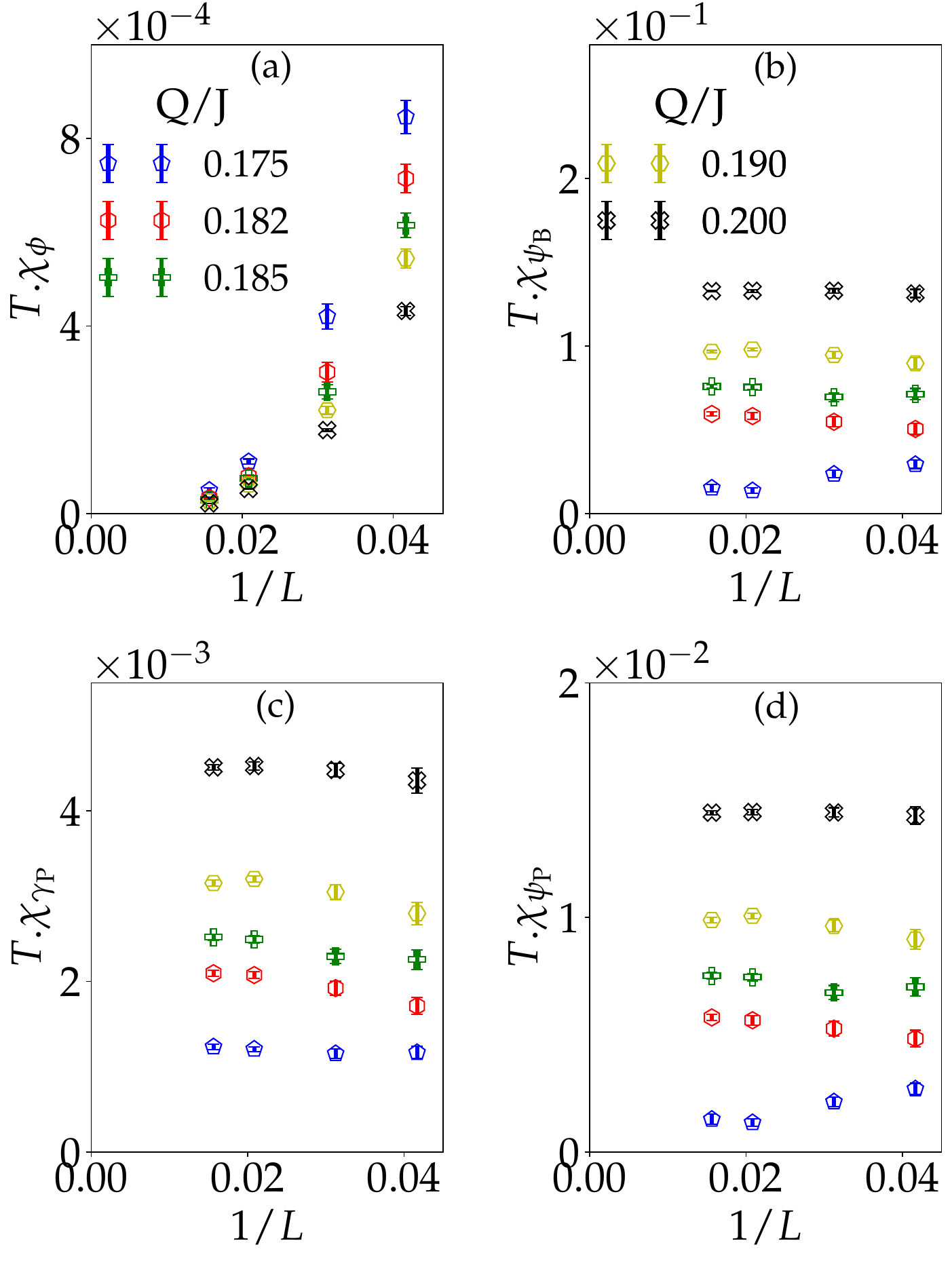}
	
	\caption{\label{fig:chi_largeQ} Susceptibility of different order parameters ($\hat{\phi}$, $\hat{\psi_{\rm B}}$, $\hat{\gamma_{\rm P}}$ and $\hat{\psi_{\rm P}}$ respectively) as defined in Eq~\ref{chi} for four different values of $g$ in the range $g\geq0.17$ at $N=12$. The susceptibility corresponding to the (a) nematic order decays down to $0$ in the thermodynamic limit where as $\chi_{\psi_{\rm B}}$, $\chi_{\gamma_{\rm P}}$ and $\chi_{\psi_{\rm P}}$ all extrapolates to non-zero values in the $L=\beta\to\infty$ limit. This provides strong evidences for the existence of p-VBS ordering at large $Q/J$. See Sec.~\ref{sec:ResultsLargerN} for details.
	}
\end{figure}

\begin{figure}
	\centering
	\includegraphics[width=\linewidth]{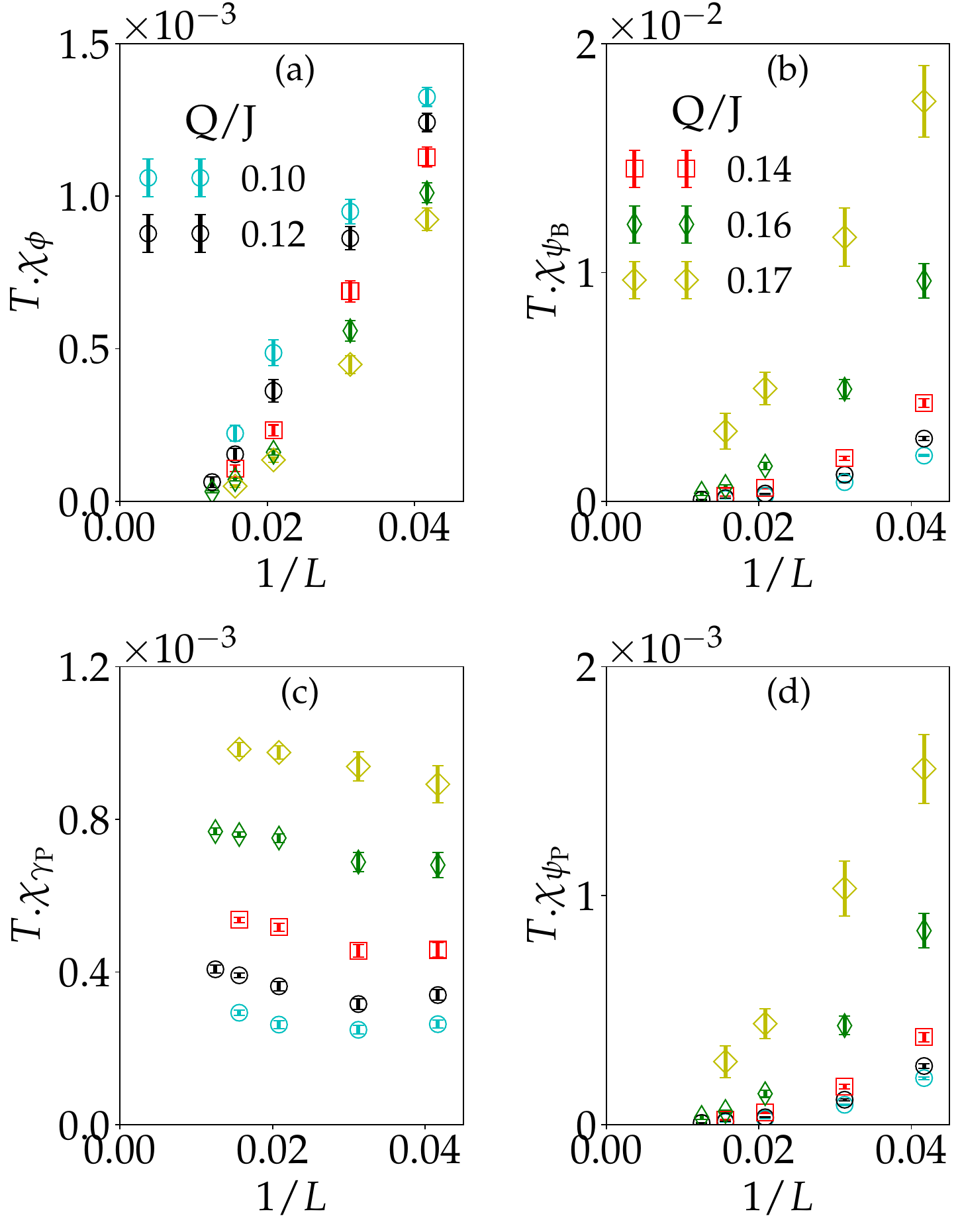}
	\caption{\label{fig:chi_intermediateQ} Susceptibility of different order parameters ($\hat{\phi}$, $\hat{\psi_{\rm B}}$, $\hat{\gamma_{\rm P}}$ and $\hat{\psi_{\rm P}}$ respectively) as defined in Eq~\ref{chi} for four different values of $g$ in the range $0.1\leq g\leq 0.17$ at $N=12$. By visually extrapolating the data points up to $1/L=0$, we can easily see that only $T\chi_{\gamma_{\rm P}}$ survives in the low temperature thermodynamic limit for this intermediate range of $Q/J$. See Sec.~\ref{sec:ResultsLargerN} for details.
	}
\end{figure}

With this in hand, we turn to the system with two minispins on each site of the square lattice. On a $A$ sublattice site, these two minispins represent a spin degree of freedom that corresponds to the symmetrized tensor product of two copies of the fundamental representation of SU($N$), while the two minispins on a $B$ sublattice site represent the symmetrized tensor product of two copies of the complex conjugate of the fundamental representation.

The antiferromagnetic exchange interaction between two of these degrees of freedom (of which one is on an $A$ sublattice site $i$  and the other on a $B$ sublattice site $j$) can now be written as
\begin{displaymath}
{\mathcal J}_{ij} = -\sum_{a,b = 1}^{2}{\mathcal S}_{ij}^{a,b} \; .
\end{displaymath}
We note that the operator ${\mathcal J}_{ij}$ acts symmetrically on the two minispins at each of the two physical sites $i$ and $j$. Also, implicit in the above is the restriction that this operator is only defined to act in the Hilbert space of our model, {\em i.e.} on states are symmetric under interchange of minispins at each physical site. Such states are unchanged by the action of the projection operator ${\mathcal P} \equiv \prod_j {\mathcal P}_j$, where ${\mathcal P}_j$ is the projector acting at site $j$ which selects states symmetric under interchange of minispin labels at that site. To include this explicitly in our definition of the Heisenberg exchange term, we may define the Heisenberg exchange Hamiltonian as 
\begin{equation}
H_{J} = J\sum_{\langle i j \rangle} {\mathcal J}_{ij} {\mathcal P} \; , 
\end{equation}
where the projection operator in front serves to remind us that this Hamiltonian is only defined for states in the physical Hilbert space of our model. Note that $H_J$ is Hermitian since ${\mathcal J}_{ij}$ commutes with ${\mathcal P}$ by virtue of the fact that it acts symmetrically on the two minispins at a physical site. This exchange term is depicted schematically in Fig.~\ref{fig:phase_diagram}.
\begin{figure}
	\centering
	{\includegraphics[width=\linewidth]{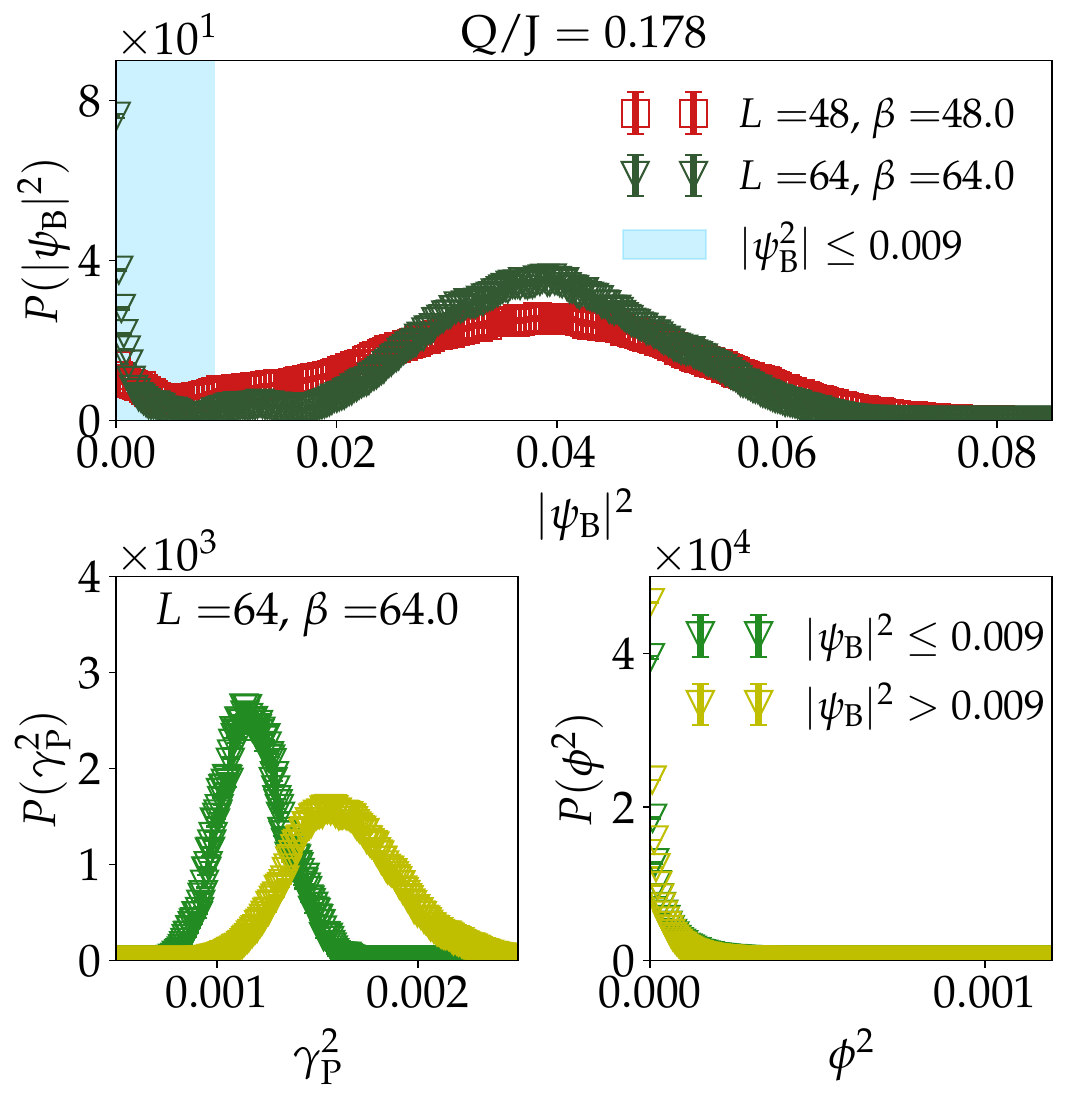}}
	\caption{\label{fig:psi_B_histogram_QbJ0.178} Top: Histogram of $\psi_\mathrm{B}^2$ at $N=12$ and $g=0.178$ for $\beta=L$ shows two prominent peaks including one non-zero peak indicating the phase coexistence corresponding to a first order transition between the ($\pi,\pi$) ordered resonating plaquette singlet phase and the pVBS phase. Bottom panels are the conditional histograms of $\gamma_\mathrm{P}^2$ and $\phi^2$ with the  condition on the values of $\psi_\mathrm{B}^2$ marked by the blue shaded region in the top figure. There are two peaks of $P(\gamma_{\rm P}^2)$ for two different values of $\gamma_{\rm P}^2\neq0$. One is because of coexisting ($\pi,\pi$) ordered plaquette singlet phase and the other is due to simultaneous Bragg peaks of the real and imaginary parts of the complex order parameters $\psi_{\rm P}$ that characterizes the pVBS phase. The nematic order histogram shows a trivial peak at $\phi^2=0$ near the transition. See Sec.~\ref{sec:ResultsLargerN} for details.
	}
\end{figure}

\begin{figure}
	\centering
	\includegraphics[width=\linewidth]{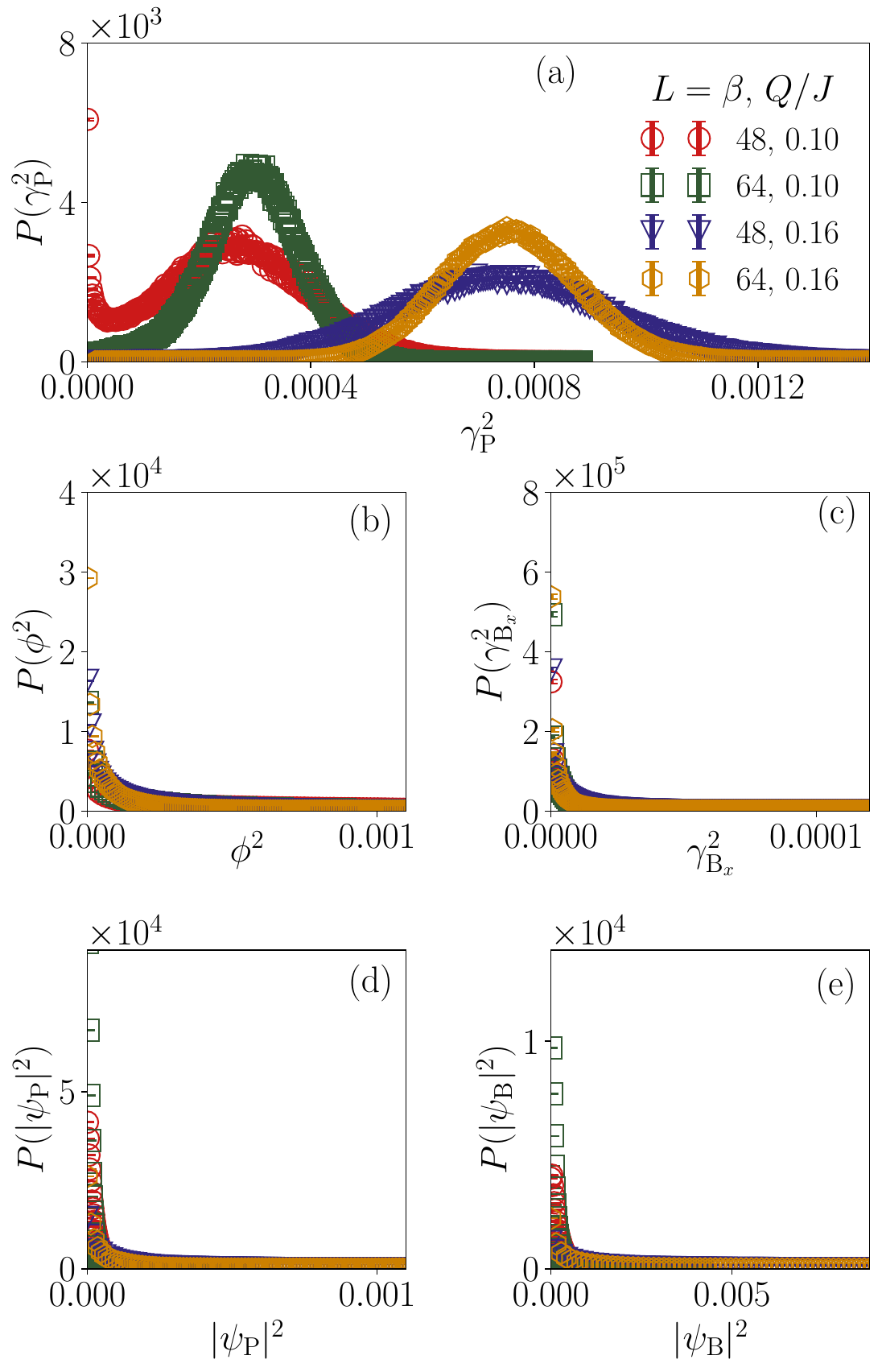}

	\caption{\label{fig:histogram_QbJ0.10and0.16} Order parameter histograms for $N=12$ at two intermediate values of coupling strength $g=0.1,0.16$ are shown for different system sizes $L=48,64$ with $\beta=L$ (low temperature limit). (a) Histograms of $\gamma_{\rm P}^2$ shows single Bragg peak at a $L$ independent but $Q/J$ dependent nonzero value of $\gamma_{\rm P}^2$ that survives in the low temperature thermodynamic limit. Histograms of $\phi^2$, $\gamma_{{\rm B}_x}^2$, $|\Psi_{\rm B}|^2$ and $|\Psi_{\rm P}|^2$ are peaked only at zero implying the absence of coresponding Haldane nematic, c-VBS and p-VBS orderings. These observations strongly suggest the existence of an unusual solid ordering of plaquette energies at wave vector ($\pi,\pi$), namely ($\pi,\pi$)-VBS in the intermediate range of coupling ($0.08<g<0.17$). See Sec.~\ref{sec:ResultsLargerN} for details.
	}
\end{figure}

The four-spin interaction term  which we expect will favour plaquette VBS order can also be written down in a very transparent manner in this language, as shown schematically in Fig.~\ref{fig:phase_diagram}. Translating this schematic depiction to an operator defined in the minispin language and acting on a single plaquette of the square lattice, we obtain:
\begin{eqnarray}
{\mathcal Q}_{ijkl} &=& -\tilde{\sum} {\mathcal S}_{ij}^{a_1,b_1}{\mathcal S}_{jk}^{b_2,c_1}{\mathcal S}_{kl}^{c_2,d_1}{\mathcal S}_{li}^{d_2,a_2} \; ,
  \label{eq:Htilde_P}
\end{eqnarray}
where $ijkl$ represents an elementary plaquette of the square lattice made up of sites $i$, $j$, $k$, and $l$ (taken, say, anticlockwise) and $\tilde{\sum}$ denotes the sum over all eight minispin indices with the proviso that only terms that satisfy $a_1 \neq a_2$, $b_1 \neq b_2$, 
$c_1 \neq c_2$, and $d_1 \neq d_2$ are included in the sum.

The corresponding plaquette term in the Hamiltonian is then given as
\begin{equation}
H_{Q} = Q \sum_{ijkl} {\mathcal Q}_{ijkl} {\mathcal P} \; ,
\end{equation}
where the ${\mathcal P}$ on the right reminds us that this term only acts on states in the physical Hilbert space of our model.
We note that ${\mathcal Q}_{ijkl}$, like ${\mathcal J}_{ij}$ discussed earlier, also acts symmetrically on all the minispin labels at each physical site, and therefore commutes with the projection operator ${\mathcal P}$. $H_{Q}$ is therefore Hermitian.
In our computational work, we study the competition between these two terms in the Hamiltonian
\begin{equation}
	\mathcal{H} = H_{J}+H_{Q} \; ,
	\label{eq:H_tot}
\end{equation}
and determine the ground state phase diagram as a function of $Q/J$ and $N$. In addition, in the p-VBS phase, we also study the finite temperature phase transition associated with the loss of p-VBS order. \begin{figure}
	\centering
	\includegraphics[width=\linewidth]{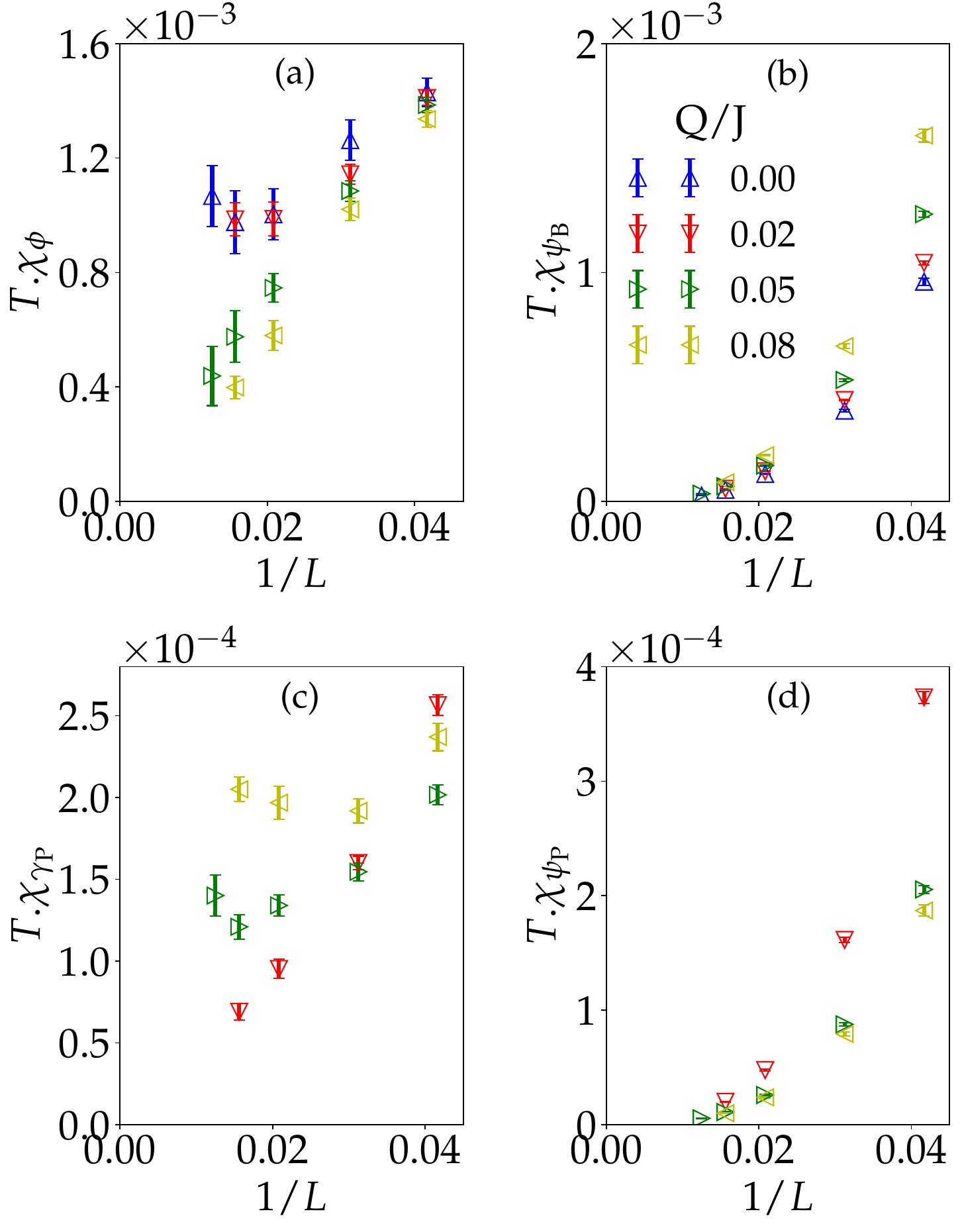}
	\caption{\label{fig:chi_smallQ} Susceptibility of different order parameters ($\hat{\phi}$, $\hat{\psi_{\rm B}}$, $\hat{\gamma_{\rm P}}$ and $\hat{\psi_{\rm P}}$ respectively) as defined in Eq~\ref{chi} for four different values of $g$ in the range $0\leq g\leq0.08$ at $N=12$. From Fig~(b) and (d) it is clear that  $\chi_{\psi_{\rm B}}$ and $\chi_{\psi_{\rm P}}$ extrapolates to $0$ at $L=\beta=\infty$ for this range of $Q/J$. At very small $Q/J$ ($\lesssim 0.05$) the nematic order starts to develop in the low temperature thermodynamic limit while the susceptibility corresponding to ($\pi,\pi$) VBS order seems to vanish at $Q/J\sim 0.02$ as is seen by visually extrapolating the corresponding data points up to $L=\beta=\infty$ in (a) and (c). See Sec.~\ref{sec:ResultsLargerN} for details.
	}
\end{figure}

\begin{figure}
	\centering
	\includegraphics[width=0.99\linewidth]{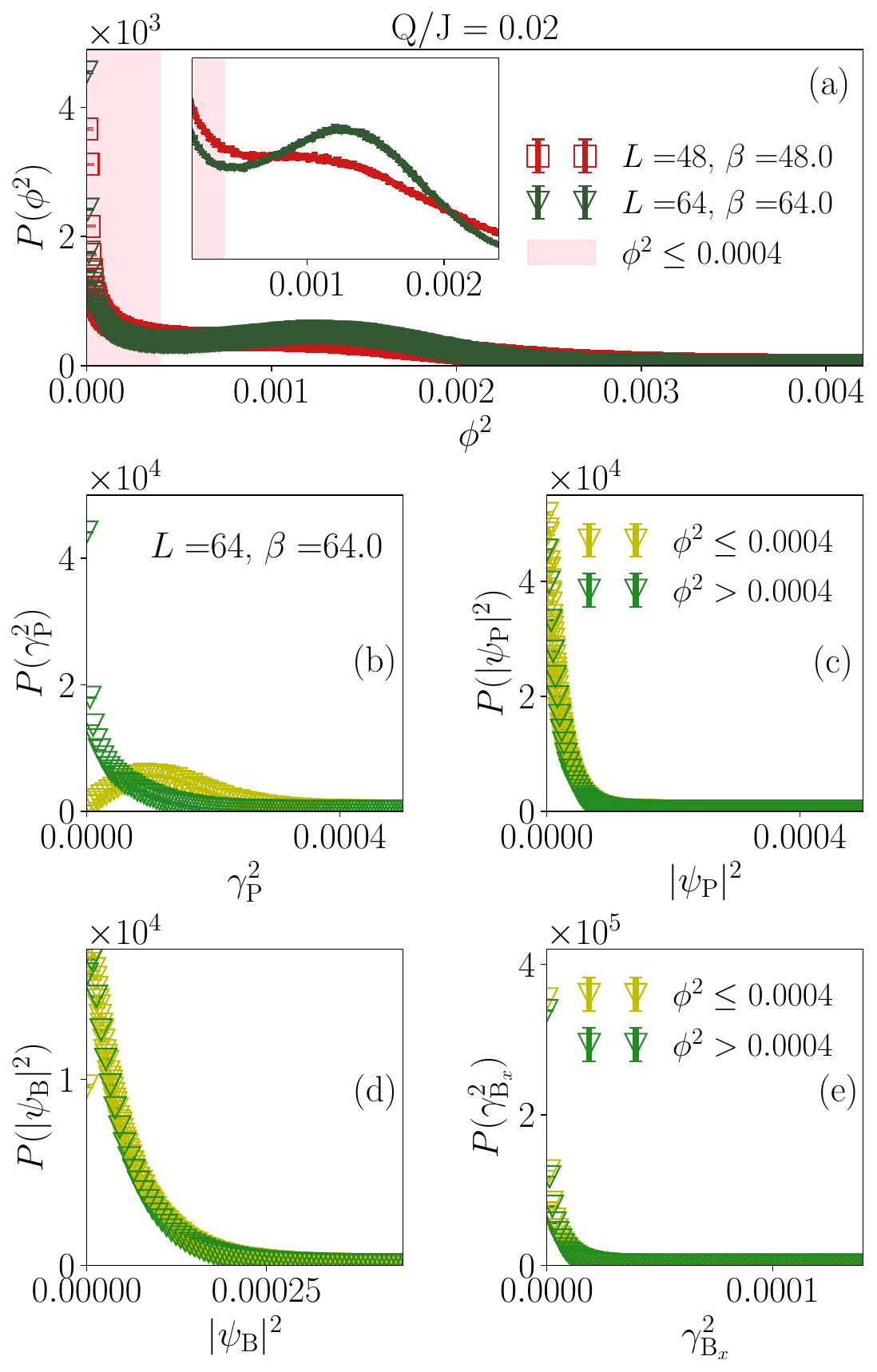}
	\caption{\label{fig:psi_histogram_QbJ0.02} (a) Histogram of the nematic order parameter $\phi^2$ at $N=12$ and $Q/J=0.02$ shows two peaks one at zero and the other at a non zero value of $\phi^2$. The second peak corresponding to a very week nematic order is clearly shown in the inset. Both the peaks getting sharper for increasing the system size as well as lowering the temperature rule out the possibility of thermal disorder and suggest some other coexisting order in the ground state. (b,c,d,e) are the conditional histograms of $\gamma_{\rm P}^2$, $|\psi_{\rm P}|^2$, $|\psi_{\rm B}|^2$ and $\gamma_{{\rm B}_x}^2$ with the condition on the values of $\phi^2$ as marked by the pink shaded region in (a). Fig~(b) suggests the coexisting phase has a Bragg peak corresponding to plaquette singlet projectors at ($\pi,\pi$) wave vector. Although there is no nontrivial Bragg peak at the columnar wave vectors ($\pi,0$) and ($0,\pi$) for (c) plaquettes and (d) bond singlet projectors and at the plaquette wave vector ($\pi,\pi$) for (e) bond singlet projectors. See Sec.~\ref{sec:ResultsLargerN} for details.
	}
\end{figure}


\section{Methods and observables}
\label{sec:MethodsandObservables}
Our computations use two different variants of the stochastic series expansion (SSE) method to study equilibrium properties of the Hamiltonian Eq.~\ref{eq:H_tot} as a function of $g \equiv Q/J$, $N$, and the inverse temperature $
\beta = 1/T$. One of them is the standard stochastic series expansion algorithm (SSE)~\cite{sandvik_aip} using the minispin representation described earlier~\cite{Todo_Kato_prl2001,Okubo_etal_prb2015,Desai_Kaul_prl2019}. When $N$ is large and/or spin correlations are short-ranged, we find that a recently-developed color-resummed version of the SSE method~\cite{Desai_Pujari_prbl2021} has distinct computational advantages over the conventional SSE method. 
  \begin{figure*}[t]
	\centering
	\includegraphics[width=0.99\textwidth]{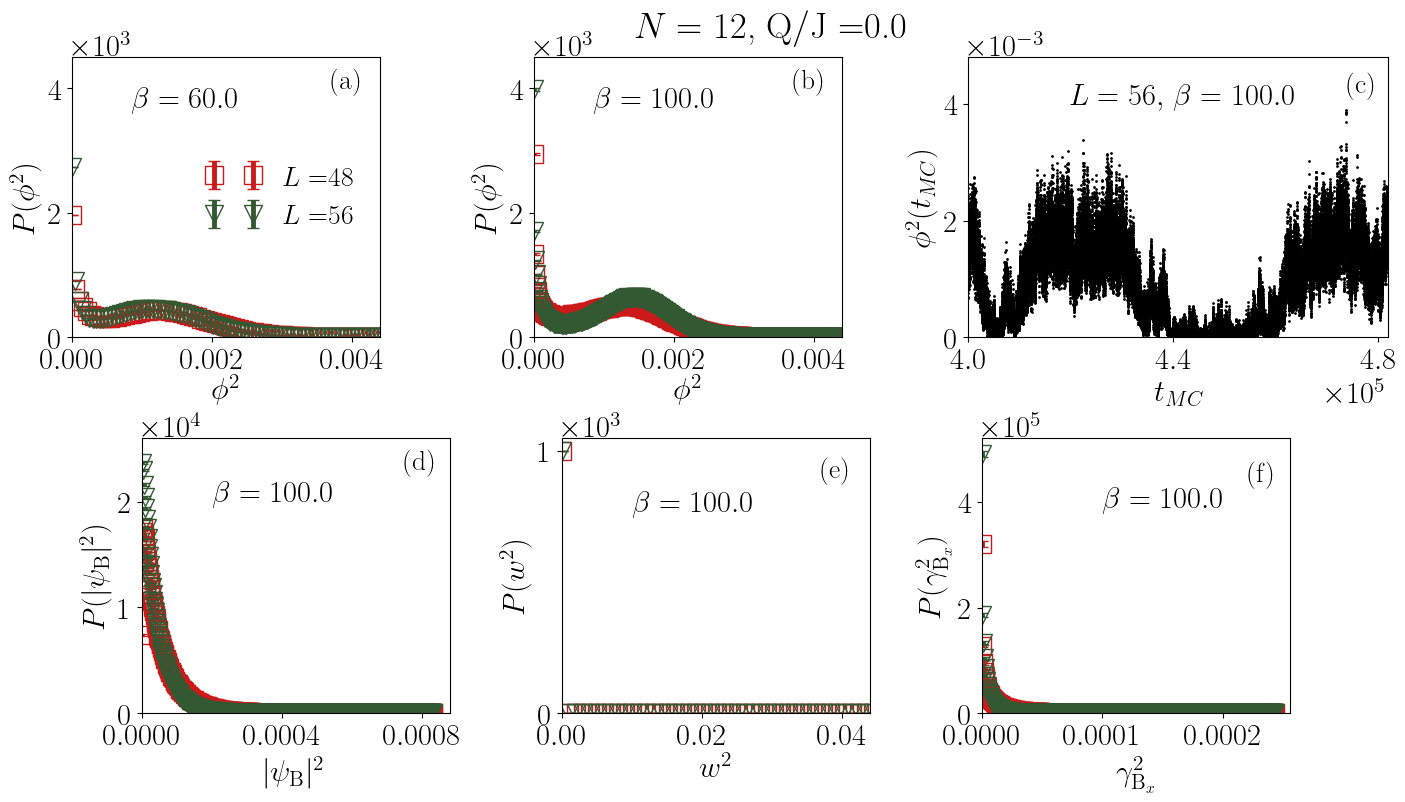}
	\caption{\label{fig:histgzero}(a,b) Histograms of the nematic order parameter $P(\phi^2)$ at $N=12$ and $g=0$ for two different values of $\beta$ shows double-peaked structure implying phase coexistence between nematic ordered phase and some other (rotational-symmetry-unbroken) phase. Both the peaks getting sharper with increasing $L$ and $\beta$ rule out the possibility of thermal disorder. (c) Monte-Carlo time series of $\phi^2$ is also seen to switch between zero and non-zero values implying the same phase coexistence. (d,e,f) Order parameter histograms corresponding to $|\psi_{\rm B}|^2$, $\rho_s$ and $\gamma_{\rm B_x}^2$ showing single peak at zero imply the absence of magnetic and any conventional c-VBS or p-VBS ordering.  See Sec.~\ref{sec:ResultsLargerN} for details.}
\end{figure*}

We have confirmed both approaches give the same results within statistical error, and have used one or the other based on the computational advantages of each in specific parameter regimes. Roughly speaking, when loops constructed in the deterministic loop update step of the algorithm are long, the original SSE approach is superior, while the color-resummed version works better in parameter regimes in which these loops are short. The former corresponds to long-range spin correlations, while the latter is typical of non-magnetic states. We have carried out MC simulation with approximately $\sim 10^5$ warm up steps for small $N$ ($N=3$). However for larger $N$, it takes longer to warm up. For example we had to take $\sim 10^6$ Monte-Carlo steps to warm up for $N=12$ while using color-resummed version of SSE. All these simulations are performed on $L\times L$ square lattices with periodic boundary conditions.

We now outline the measurements we have performed to characterize different phases and transitions among them. The  SU(N) symmetric analog of long range antiferromagnetic order of SU(2) symmetric $S=1$ antiferromagnets is long range correlation in the color  (SU(N) index) variable. This is characterized by the correlation function 
\begin{equation}
	\mathcal{C}_{m}(\mathbf{r}) = \frac{1}{N \times N_{s}}\sum_{\alpha,\mathbf{\rm \mathbf{x}}}\langle\hat{J}^{\alpha}_{\alpha}(\mathbf{\rm \mathbf{x}})\hat{J}^{\alpha}_{\alpha}(\mathbf{\rm \mathbf{x}}+\mathbf{r})\rangle,
\end{equation}
where $\hat{J}^{\alpha}_{\alpha}=|\alpha\rangle\langle \alpha|-1/N$ is a diagonal generator of SU($N$)\footnote[1]{$1/N$ is subtracted to make the operator traceless} and $N_s=L^2$ is the number of sites in a $L\times L$ square lattice. In the magnetic phase, the Fourier transform,

\begin{equation}
	\tilde{\mathcal{C}}_{m}(\mathbf{k})=\sum_{\mathbf{r}} e^{i \mathbf{k}.\mathbf{r}} \mathcal{C}_{m}(\mathbf{r})
\end{equation}
 has a Bragg peak at $\mathbf{k_{0}}=(0,0)$ and the peak height is proportional to the square of the magnetization.

In addition we also define the dimensionless ratio 
\begin{equation}
	\mathcal{R}_{m} = 1-\frac{\tilde{\mathcal{C}}_{m}(\mathbf{k_0^{'}})}{\tilde{\mathcal{C}}_{m}(\mathbf{k_0})}.
	\label{eq:ratio}
\end{equation}
Here $\mathbf{k_0}$ is the ordering momentum and $\mathbf{k_0^{'}}$  is the momentum closest to the ordering momentum. 

The spin stiffness is another quantity used to detect the magnetic phase. It can be measured by adding a twist in the boundary condition and measuring the second derivative of the energy of the system with respect to the twist angle:
 \begin{equation}
  \rho_s=\frac{\partial^2 E(\theta)}{\partial \theta^2}\bigg|_{\theta=0} \\
  \label{stiff}
 \end{equation}
Here E($\theta$) is the energy of the system when you add a twist of $\theta$ in the boundary condition in either the $x$ or the $y$ direction. In the QMC, this quantity is related to the winding number of loops in the direction that the twist has been added:
\begin{equation}
  \rho_s= \frac{1}{N}\sum^N_{\alpha=1}\frac{ \overline{W_{\alpha}^2}}{\beta}\equiv\overline{w^2}\\
  \label{eq:winding}
 \end{equation}
  where $\beta$ is the inverse temperature. The winding number of loops of each color $\alpha$ is measured and the average is taken over all the colors. Here $w^2$ is the Monte-Carlo estimator for the stiffness.
\begin{figure}
	\centering
	\subfigure[]{\includegraphics[width=\linewidth]{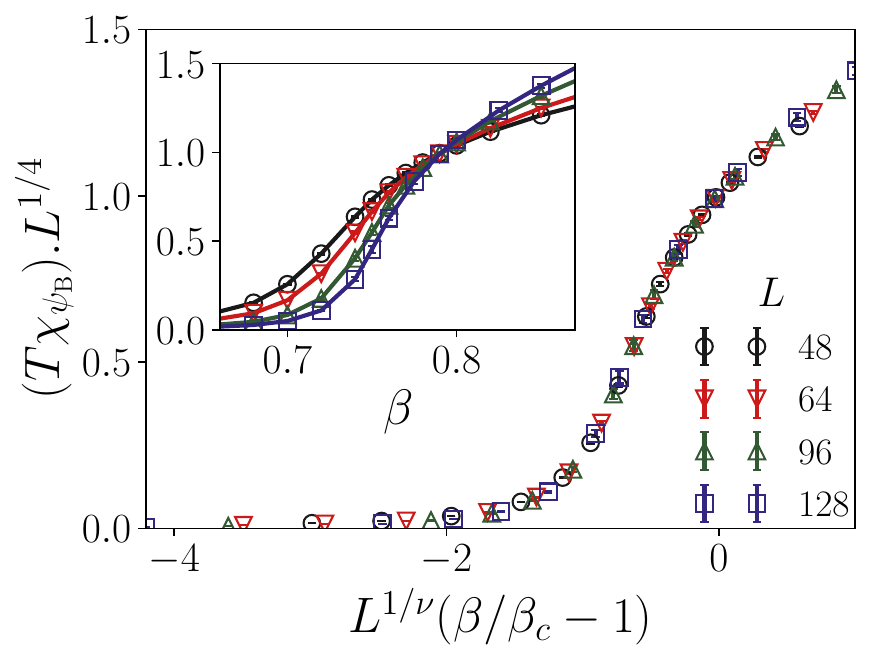}}
	\subfigure[]{\includegraphics[width=\linewidth]{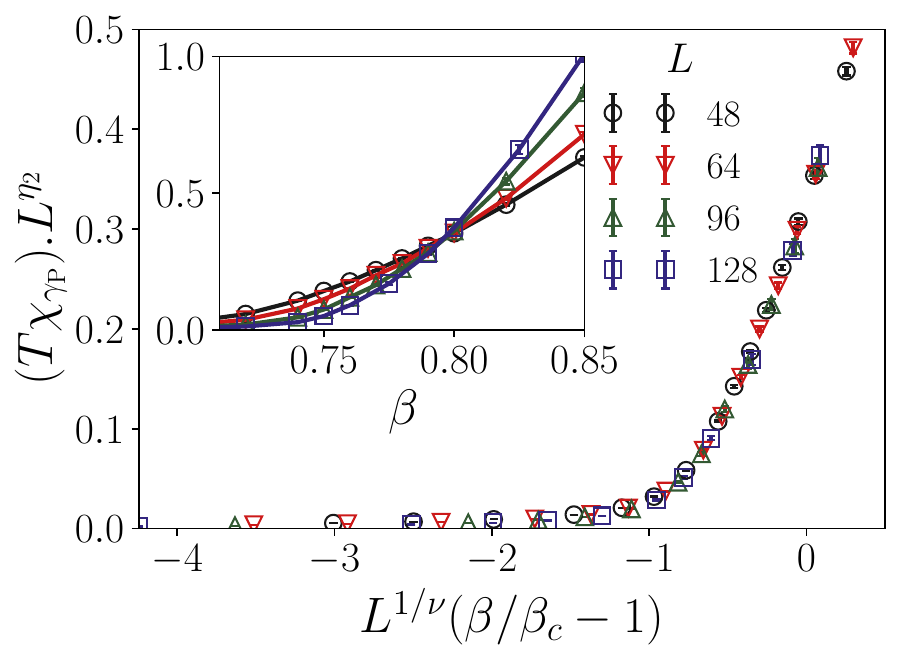}} 
	\caption{\label{fig:N3_AT_phi2,Qpipi} (a) Inset: $|\psi_{\rm B}|^{2}.L^{1/4}$ for $N=3$ and $g=3.0$ plotted as a function of $\beta$ for different system sizes ($L$). The curves corresponding to different system sizes show a clear crossing allowing us to estimate $\beta_c=0.792(5)$. (a) Scaling collapse of $|\psi_{\rm B}|^{2}.L^{1/4}$ for various system sizes close and below to the critical point with the correlation length exponent $\nu=1.85(10)$.
		(b) Inset: $\gamma_{\rm P}^{2}.L^{\eta_2}$ for $N=3$ and $g=3.0$ plotted as a function of $\beta$ for various system sizes cross at the estimated $\beta_c=0.792(5)$ when $\eta_2$ is set to $0.76(6)$. (b) Scaling collapse of $\gamma_{\rm P}^{2}.L^{\eta_2}$ for various $L$ close and below to the critical point using $\nu=1.85$ and $\beta_c=0.795$ and $\eta_2=0.73$ belonging to the aforementioned estimated range. See Sec.~\ref{sec:AshkinTeller} for details.}
\end{figure}

In order to identify the HN phase, we construct the following order parameter:

\begin{equation}
	\hat{\phi} = \frac{1}{N_s}\sum_{\mathbf{r}}\hat{\phi}(\mathbf{r}),
\end{equation}
where $\hat{\phi}(\mathbf{r})$ is the local nematic order parameter defined as
\begin{equation}
	\hat{\phi}(\mathbf{r}) = \frac{1}{2}\left(\mathcal{J}_{\rm \mathbf{r},\mathbf{r+\hat{x}}}+\mathcal{J}_{\rm \mathbf{r},\mathbf{r-\hat{x}}}-\mathcal{J}_{\rm \mathbf{r},\mathbf{r+\hat{y}}}-\mathcal{J}_{\rm \mathbf{r},\mathbf{r-\hat{y}}} \right).
	\label{eq:phi}
\end{equation}
Here $\mathcal{J}_{\rm \mathbf{r},\mathbf{r\pm\hat{x}(\hat{y})}}$ is the projection operator (defined in the previous section) between site at $\mathbf{r}$ and it's nearest neighbor site along $\pm\mathbf{\hat{x}(\hat{y})}$ direction.

The pVBS phase on the other hand breaks symmetry of lattice translation along both $x$ and $y$ direction. We construct the following two complex order parameter: 
\begin{equation}
	\hat{\psi}_{\rm B} = \frac{1}{N_s} \sum_{\mathbf{r}}\hat{\psi}_{\rm B}(\mathbf{r}), \hspace{0.5cm} \hat{\psi}_{\rm P} = \frac{1}{N_s} \sum_{\mathbf{r}}\hat{\psi}_{\rm P}(\mathbf{r})
	\label{eq:psi}
\end{equation}
using singlet projectors of bonds and plaquettes respectively. $\hat{\psi}_{\rm B}(\mathbf{r})$ and $\hat{\psi}_{\rm P}(\mathbf{r})$ correspond to local complex order parameters characterizing the VBS order defined as:
\begin{align}
	\hat{\psi}_{\rm B}(\mathbf{r}) &= \frac{(-1)^{\mathbf{r}_x}}{2}\left(\mathcal{J}_{\rm \mathbf{r},\mathbf{r+\hat{x}}}-\mathcal{J}_{\rm \mathbf{r},\mathbf{r-\hat{x}}}\right) \nonumber\\
	&+i\frac{(-1)^{\mathbf{r}_y}}{2}\left(\mathcal{J}_{\rm \mathbf{r},\mathbf{r+\hat{y}}}-\mathcal{J}_{\rm \mathbf{r},\mathbf{r-\hat{y}}}\right),\\
	\hat{\psi}_{\rm P}(\mathbf{r}) &= (-1)^{\mathbf{r}_x}\mathcal{Q}(\mathbf{r})+i(-1)^{\mathbf{r}_y}\mathcal{Q}(\mathbf{r}).
	\label{eq:psi(r)}
\end{align}
Here $\mathcal{Q}(\mathbf{r})$ is the same plaquette singlet projector as defined in Eq.~\ref{eq:Htilde_P} where we have omitted the site indices of four corners and used position of the left bottom corner of a plaquette $\mathbf{r}$ to specify the position of $\mathcal{Q}$. 
Here $(\mathbf{r}_x,\mathbf{r}_y)$ are the coordinates of the lattice vector $\mathbf{r}$.
We also define another set of operator ($\hat{\gamma}_{\rm B (P)} = \frac{1}{N_s}\sum_{\mathbf{r}}\hat{\gamma}_{\rm B (P)}(\mathbf{r})$) that correspond to Fourier transform of bond and plaquette singlet projectors at ($\pi,\pi$) wave vector. These are namely
\begin{align}
	&\hat{\gamma}_{{\rm B}_{x(y)}}(\mathbf{r}) = (-1)^{\mathbf{r}_x+\mathbf{r}_y}\mathcal{J}_{\mathbf{r},\mathbf{r+\hat{x}(\hat{y})}},\\
	&\hat{\gamma}_{\rm P}(\mathbf{r}) = (-1)^{\mathbf{r}_x+\mathbf{r}_y}\mathcal{Q}(\mathbf{r}).
	\label{eq:gamma(r)}
\end{align}

For a general order parameter $\hat{\mathcal{O}}$, the quantum mechanical expectation,
\begin{equation}
	\langle\hat{\mathcal{O}}\rangle =\overline{\mathcal{O}}
\end{equation}
where $\overline{\mathcal{O}}$ is the Monte-Carlo average of the estimator $\mathcal{O}$ for the operator $\hat{\mathcal{O}}$.

The columnar VBS order breaks symmetry of lattice  translation in any one of $x,y$ direction. That leads to the complex order parameters having phase angle any one of $0,\pi/2,\pi,3\pi/2$.  In contrast plaquette VBS order breaks translation-symmetry in both $x,y$ direction leading the complex order parameters to have phase angle any one of $\pi/4,3\pi/4,5\pi/4,7\pi/4$. Thus one can distinguish between pVBS and cVBS phase by measuring the joint histograms of estimators for the real and imaginary part of $\hat{\psi}_{\rm B}$.
To locate the phase transition corresponding to the onset of this p-VBS order, we also find it useful to measure the equal time structure factor $\tilde{\mathcal{C}}_{\psi_B}$ and use it to construct the dimensionless ratio $\mathcal{R}_{\psi_B}$ in a manner entirely analogous to the definition of ${\mathcal R}_{m}$.



Additionally we measure the static susceptibility corresponding to the order parameters defined above. For a general order parameter $\mathcal{\hat{O}}=\sum_{\mathbf{r}}\hat{\mathcal{O}}(\mathbf{r})/N_s$ the static susceptibility (scaled by volume) is defined as:

\begin{align}
	\chi_{\mathcal{O}} &= \int_{0}^{\beta}d\tau\langle e^{\tau\mathcal{H}}\mathcal{\hat{O}} e^{-\tau\mathcal{H}}\mathcal{\hat{O}^{*}}\rangle,\\
	\overline{\mathcal{O}^2} &= \chi_{\mathcal{O}}/\beta.
	\label{chi}
\end{align}
Here $\mathcal{O}^2$ stands for the Monte-Carlo estimator for the static correlator of $\hat{\mathcal{O}}$ and the bar stands for Monte-Carlo average.

\section{Small $N$ results: Color ordered (Neel) to p-VBS transitions}
\label{sec:ResultsSmallerN}

We begin by studying the phase diagram at smaller values of $N$, up to $N=9$.
At small values of $Q/J$, one expects color order which is the SU(N) analog of antiferromagnetism in the $S=1$ square lattice antiferromagnets.  As described in the Introduction, we expect however that large values of $Q/J$ favor a state that minimizes the expectation value of the four spin plaquette interaction term. An intuitively appealing and simple variational wavefunction for such a state can be constructed from a maximally packed non-overlapping pattern ({\em i.e.} not sharing an edge or a vertex) of ``occupied'' plaquettes as shown in Fig.~\ref{fig:quantum_phases}, where each ``occupied'' plaquette denotes the singlet state constructed from the four spins on its vertices. \begin{figure}
	\centering
	\subfigure[]{\includegraphics[width=\linewidth]{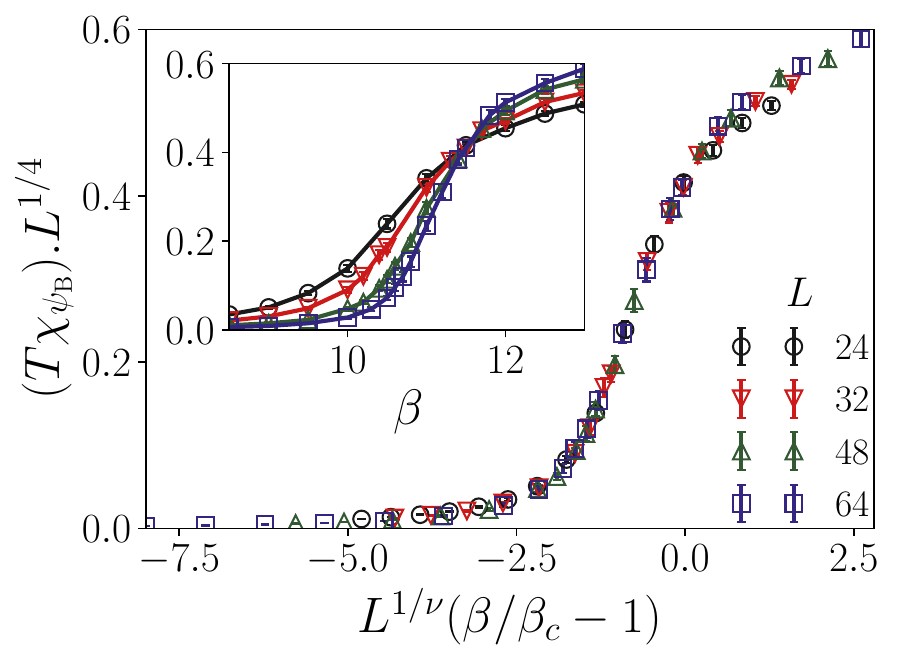}}
	\subfigure[]{\includegraphics[width=\linewidth]{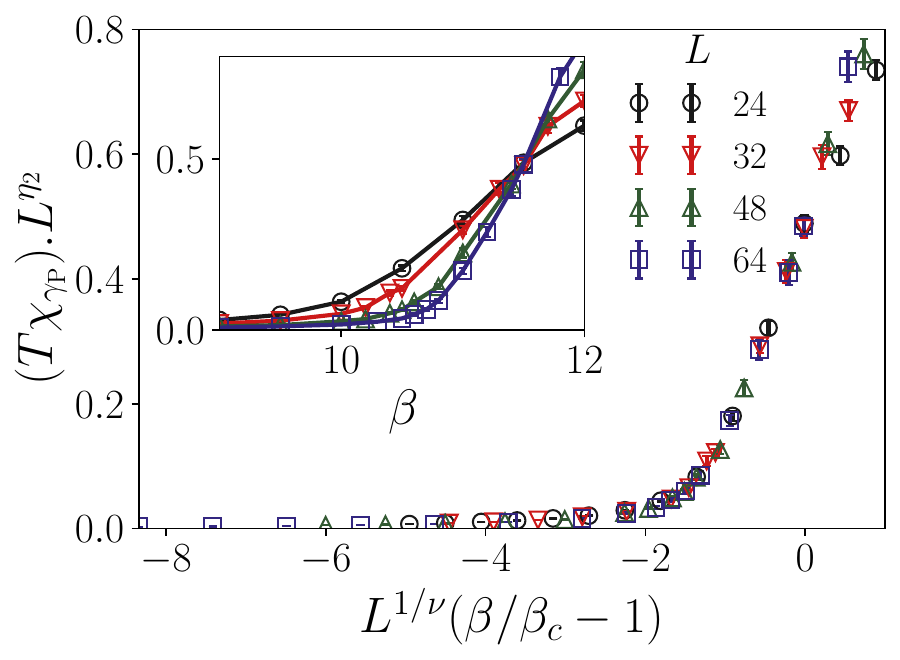}} 
	\caption{\label{fig:N12_AT_phi2,Qpipi} (a) Inset: $|\psi_{\rm B}|^{2}.L^{1/4}$ for $N=3$ and $g=3.0$ plotted as a function of $\beta$ for different system sizes ($L$). The curves corresponding to different system sizes show a clear crossing allowing us to estimate $\beta_c=11.525$. (a) Scaling collapse of $|\psi_{\rm B}|^{2}.L^{1/4}$ for various system sizes close and below to the critical point with the correlation length exponent $\nu=1.37(5)$.
		(b) Inset: $\gamma_{\rm P}^{2}.L^{\eta_2}$ for $N=12$ and $g=0.25$ plotted as a function of $\beta$ for various system sizes cross at the estimated $\beta_c=11.525$ when $\eta_2=0.60(5)$. (b) Scaling collapse of $\gamma_{\rm P}^{2}.L^{\eta_2}$ for various $L$ close and below to the critical point using $\beta_c=11.51$, $\nu=1.36$ and $\eta_2=0.632$ belonging to the aforementioned estimated range. See Sec.~\ref{sec:AshkinTeller} for details.}
\end{figure}

Since this variational state necessarily breaks lattice translation symmetry simultaneously in both the $x$ and $y$ directions, we expect a valence bond solid ordered state at large $Q/J$. This valence bond solid state has plaquette order, and we dub it the plaquette VBS (p-VBS) state. As mentioned earlier, it is expected to show simultaneous Bragg peaks at wavevector $(\pi, 0)$ in the static structure factor of the $x$ bond energies (singlet projector on bonds) and at wavevector $(0,\pi)$ in the static structure factor of the $y$ bond energies. 
In addition, as a subdominant feature, it is also expected to show a Bragg peak at wavevector $(\pi, \pi)$ in the static structure factor of the plaquette energies (four-spin interaction operator on plaquettes). However, it is expected to preserve symmetry of rotations about a plaquette center.

This intuition motivates our detailed study of the ground state phase diagram as a function of $Q/J$ for a range of values of $N$.  To test for the presence of color order, we monitor the stiffness $\rho_s$ that is the SU(N) analog of the antiferromagnetic spin stiffness of SU(2) antiferromagnets. In addition, we measure the the equal time structure factor $\tilde{\mathcal{C}}_{m}$ of the SU(N) color quantum number (which is the analog of the equal time spin structure factor of SU(2) antiferromagnets) and the static color susceptibility. To test for the presence of p-VBS order, we measure the corresponding p-VBS parameter susceptibility $\chi_{\psi_B}$, as well as the corresponding equal time structure factor $\tilde{\mathcal{C}}_{\psi_B}$.  In addition, we also construct and monitor the ratios $\mathcal{R}_{m}$ and $\mathcal{R}_{\psi_B}$ of the equal time color structure factor and the equal time structure factor of the p-VBS order parameter $\hat{\psi}_B$ respectively. These dimensionless ratios are useful as a means of identifying the location of the phase transition between the color ordered state and the p-VBS state.

We find that color order persists in the $N=2$ case even when $J=0$ and $Q>0$. However, for $N \geq 3$ we do find that color order gives way to p-VBS order above a threshold value of $Q/J$, exactly in accordance with the intuition outlined above. As a representative example, we display in Fig.~\ref{fig:corrratios_N3} our small size data for the structure factor ratios ${\mathcal R}_{m}$ and ${\mathcal R}_{\psi_B}$ for the $N=3$ case. It is clear from this data that a non-magnetic p-VBS ordered state is established for $Q/J \gtrsim 1.83$. From this data, it also appears that the transition is first order in nature, since the dimensionless ratio curves steepen and shift extremely rapidly with increasing size instead of having a well-defined crossing point that would signify a second-order phase transition. 

To explore this further, we have monitored the low temperature Monte Carlo time series and histograms of the SSE estimator $w^2$ of the spin stiffness $\rho_s$, as well as the SSE estimator $E_{\psi_B}$ of the equilibrium expectation value $\langle \hat{\psi}_B\rangle$. These results are displayed in Fig.~\ref{fig:hist_transition_N3}.
 for $N=3$ in the vicinity of the transition. They show clear evidence of the first order nature of the transition: First, the histogram of $w^2$ has a pronounced two peak structure. Second, the Monte Carlo time series of $w^2$ shows characteristic ``stickiness'', jumping intermittently between two well-defined plateaus. Third, the histogram of $E_{\psi_B}$ shows two sets of peaks, one at $E_{\psi_B} = 0$ and the other four at a nonzero value of $|E_{\psi_B}|$. And finally, this peak structure is seen to sharpen noticeably with increasing size $L$. 
The nature of the transition at larger values of $N$ in this range ({\em i.e.} $ N \lesssim 9$) is similar. This is clear from the analogous histogram data for $N=8$ displayed in Fig.~\ref{fig:hist_transition_N8}. For this figure, we see that the first order nature of the color order to p-VBS order transition is actually much more pronounced at $N=8$.
 
Finally, we note that the occurrence of a first order transition is not surprising here, given that the spins are SU(N) analogs of $S=1$ objects. While the possibility of a second order transition between the N\'eel and VBS phase has been discussed earlier for $S=1/2$, the original theory of deconfined criticality does not predict the same phenomenon for $S=1$ or other integer spins~\cite{senthil2004:science,Senthil2004}. Parenthetically, we note however that the possibility of such a second order transition has been explored in a field-theoretical analysis of the N\'eel - Haldane Nematic transition for S=1~\cite{wang2015:s1}. However, in this work we find a conventional first order phase transition when the nonmagnetic phase has plaquette VBS order.  Our results should also be contrasted with those of another recent study of a S=1 model that has been shown to host a "pseudo-critical" or very weakly first order transition between Neel and columnar VBS phases~\cite{Vijigiri_etal_arxiv}.

\section{Results for $N > 9$: Competition between Haldane nematic, p-VBS, and $(\pi,\pi)$-VBS order }
\label{sec:ResultsLargerN}
We have also studied the phase diagram for larger values of $N$, {\em i.e.} $N > 9$. For large enough $Q/J$, we again find a non-magnetic p-VBS state entirely analogous to the corresponding state studied earlier at smaller values of $N$. However, upon lowering the value of $Q/J$, we find an interesting sequence of non-magnetic VBS phases that point to the delicate energetic balance between various kinds of valence bond solid orders.

To study this in some detail, we focus on the representative case of $N=12$ and monitor  i) the VBS order parameters $\hat{\psi}_B$ and $\hat{\psi}_P$ that are sensitive to bond and plaquette order respectively at the wavevectors $(\pi, 0)$ and $(0,\pi)$ , ii) the Haldane nematic order parameter $\hat{\phi}$ that is sensitive to lattice rotation symmetry breaking in the bond energies, and iii) the $(\pi,\pi)$ order parameter $\hat{\gamma}_P$ that is sensitive to ordering of the plaquette energies at wavevector $(\pi,\pi)$.

In Fig.~\ref{fig:chi_largeQ}, we display the results for the $L$ dependence of the static susceptibilities at inverse temperature $\beta = L$ for  values of $Q/J$ in the large $Q/J$ regime, with $Q/J  \gtrsim 0.18$. We see that $T\chi_\phi$ goes to zero rapidly with increasing $L$, while $T \chi_{\psi_B}$ and $T \chi_{\psi_P}$ go to an nonzero limit as $L$ is increased. This is what one expects of the p-VBS phase. From the $L$ dependence of $T \chi_{\gamma_P}$, we also see that p-VBS order is associated with subdominant long range order of the plaquette energies at wavevector $(\pi, \pi)$. This is exactly what one expects from the simple variational state depicted in Fig.~\ref{fig:quantum_phases}.

These results for the various valence bond order parameter susceptibilities should be contrasted with the corresponding results at $\beta = L$ for intermediate values of $Q/J$, in the range $0.1 \lesssim Q/J \lesssim 0.17$. These are displayed in Fig.~\ref{fig:chi_intermediateQ}. In this regime, we see that $T \chi_{\gamma_P}$ tends to a nonzero large $L$ limit, while $T \chi_{\psi_B}$, $T \chi_{\psi_P}$, and  $T\chi_\phi$ all tend to zero.
This is an unambiguous indication of the presence of a phase in which the plaquette energies are ordered at wavevector $(\pi, \pi)$, but there is no ordering of the bond or plaquette energies at wavevectors $(\pi, 0)$ and $(0, \pi)$, nor is there any sign of Haldane nematic order.

The nature of the transition from the large $Q/J$ p-VBS phase to this intermediate $Q/J$ $(\pi, \pi)$-VBS phase is best understood by monitoring the histograms of the estimators $|\psi_{\rm B}|^2$, $\gamma^2_{\rm P}$, and $\phi^2$ of the corresponding scaled susceptibilities $T\chi_{\psi_{\rm B}}$, $T\chi_{\gamma_{\rm P}}$, and $T\chi_\phi$  in the vicinity of the transition. This is shown in Fig.~\ref{fig:psi_B_histogram_QbJ0.178}. We see that there is a clear double peak structure in the histogram of  $|\psi_{\rm B}|^2$, with one peak at zero and another at an $L$ independent nonzero value. 

When one filters the data based on whether a configuration belongs to the peak at zero or the nonzero peak, we see quite clearly that the histogram of $\phi^2$ remains peaked at zero for both kinds of configurations. However, the two kinds of configurations correspond to peaks at two different nonzero values of $\gamma^2_{\rm P}$, with the peak at a higher value corresponding to configurations which belong to the nonzero peak of $|\psi_{\rm B}|^2$. We have already seen that p-VBS order is also associated with subdominant order of the plaquette energies at wavevector $(\pi, \pi)$. 

This is consistent with the fact that the  peak at a higher value of $\gamma^2_{\rm P}$ corresponds to nonzero values of $|\psi_{\rm B}|^2$. The interpretation of peak at the lower value of $\gamma^2_{\rm P}$ is then that it corresponds to configurations which have $(\pi, \pi)$ order without any p-VBS order. This is exactly what we have already seen in the intermediate $Q/J$ phase in Fig.~\ref{fig:chi_intermediateQ}. This interpretation is further corroborated by the histograms of these estimators in the intermediate $Q/J$ regime. These are shown in Fig.~\ref{fig:histogram_QbJ0.10and0.16}. From these histograms, we again conclude that the intermediate phase is associated with $(\pi, \pi)$ ordering of the plaquette energies without any p-VBS order or Haldane nematic order.
Another point to note is that the bond energies do {\em not} show any ordering at the $(\pi, \pi)$ wavevector although the plaquette energies are ordered at this wavevector.

Next we turn to the low $Q/J$ regime. In Fig.~\ref{fig:chi_smallQ}, we display the susceptibilities corresponding to p-VBS, $(\pi, \pi)$-VBS, and Haldane nematic orders for inverse temperatature $\beta = L$ and a sequence of sizes. In this low $Q/J$ regime,  $T \chi_{\psi_B}$ and $T \chi_{\psi_P}$ both tend to zero with increasing $L$. For $Q/J \gtrsim 0.05$, $T\chi_{\phi}$ also tends to zero as $L$ is increased, but when $Q/J \lesssim 0.02$, it tends to a nonzero value in the large $L$ limit.  On the other hand, $T \chi_{\gamma_{\rm P}}$ tends to a nonzero limit as $L$ is increased when $Q/J \gtrsim 0.05$. The large $L$ behavior of $T \chi_{\gamma_{\rm P}}$ for $Q/J \lesssim 0.02$ is less clear, in the sense that it is not clear if it goes to a small nonzero value or not. This is a difficult question to resolve, since the number of plaquette operators in the SSE operator string itself goes to zero as $Q/J$ goes to zero.

To shed more light on this, we turn to an examination of the histograms of the estimators $|\psi_{\rm B}|^2$, $|\psi_{\rm P}|^2$, $\gamma^2_{\rm P}$, $\gamma^2_{\rm B}$, and $\phi^2$ of the corresponding scaled susceptibilities at $Q/J = 0.02$ for range of $L$ and inverse temperature $\beta =L$. These are shown in Fig.~\ref{fig:psi_histogram_QbJ0.02}. We see a marked double peak structure in the histogram of $\phi^2$, with this structure becoming more accentuated at larger values of $\beta = L$. One of these peaks is at $\phi^2 = 0$, while the other is at a $L$ independent value of $\phi^2$. Thus, this low $Q/J$ regime seems to feature a competition between the Haldane nematic state and another state. The nature of this competition is of course quite clear when one examines the histogram of $\gamma_{\rm P}^2$ obtained from configurations that are sorted according to the value of $\phi^2$. We see that configurations belonging to the nonzero peak of the histogram of $\phi^2$ contribute to a peak at zero in the histogram of $\gamma_{\rm P}^2$. On the other hand, configurations that belong to the peak at zero in the histogram of $\phi^2$ contribute to a peak at a nonzero value in the histogram of  $\gamma_{\rm P}^2$. No such bimodality or anticorrelation is seen in the histograms of $|\psi_{\rm B}|^2$, $|\psi_{\rm P}|^2$, and $\gamma^2_{\rm B}$. Thus we conclude that there is a first order transition somewhere in the interval $Q/J \in (0.02, 0.07)$ at which the $(\pi, \pi)$-VBS order is lost and Haldane nematic order established at lower values of $Q/J$. 

However, the fact that we continue to see an anticorrelated double peak structure in the histograms of $\gamma_{\rm P}^2$ and $\phi^2$ even at the lowest values of $Q/J$ at which we are able to measure $T\chi_{\gamma_{\rm P}}$ strongly suggests that the $(\pi, \pi)$-VBS ordered state survives as a metastable state even in the $Q/J \rightarrow 0$ limit. Further (albeit indirect) evidence in favor of this conclusion comes from an examination of the histograms of $|\psi_{\rm B}|^2$,  $\gamma^2_{\rm B}$, $\phi^2$ and $w^2$ at $Q/J = 0$, and an examination of the Monte Carlo time series of $\phi^2$. These are shown in Fig.~\ref{fig:histgzero}From this data, we see that there is a two peak structure in the histogram of $\phi^2$, with one peak at zero and another at a nonzero value of $\phi^2$. Moreover, this two peak structure {\em sharpens} as the temperature is lowered to correspond to $\beta = 100$, thereby ruling out the possibility that this is the effect of thermal fluctuations. The same bimodality is also reflected in the corresponding time series. However, no corresponding bimodality or two peak structure is visible in the histograms of  $|\psi_{\rm B}|^2$,  $\gamma^2_{\rm B}$,  and $w^2$. This rules out the possibility that competing magnetic order or p-VBS order is responsible for the observed low-temperature two peak structure in the histogram of $\phi^2$. When viewed in the light of the previously displayed evidence for a metastable $(\pi, \pi)$-VBS ordered state at very low nonzero $Q/J$, this strongly suggests that the Haldane nematic state is very close in energy to a metastable $(\pi, \pi)$-VBS ordered state even at $Q/J = 0$.


\section{Ashkin-Teller criticality}
\label{sec:AshkinTeller}

The p-VBS order at wavevectors $(\pi, 0)$ and $(0, \pi)$ is described in terms of the complex order parameter field $\hat{\psi}_{\rm B}$. From the symmetries of this local order parameter under lattice transformations, we see that a good description of the finite temperature melting of this order would be provided by an $xy$ model with four-fold anisotropy, or, equivalently, an Ashkin-Teller model comprising two coupled Ising degrees of freedom $\sigma$ and $\tau$~\cite{Ashkin_Teller_1943}. The symmetries of the problem dictate that we may identify $E_{\psi_{\rm B}} \sim \sigma + i \tau$. Note that this is different from the identification we would make if the ordering was of the columnar VBS type. In this latter case, one would have written $E_{\psi_{\rm B}} \sim (\sigma+\tau) + i (\sigma -\tau)$~\cite{Ramola_Damle_Dhar_prl2015}. 

These symmetry based considerations strongly suggest that the thermal melting of p-VBS order will be in the Ashkin-Teller universality class~\cite{Kadanoff_Brown_Ann_of_phys1979,Kadanoff_jphysA1978}. In such Ashkin-Teller transitions, the anomalous exponent $\eta$ corresponding the correlation function of the four-fold symmetry breaking field is fixed to $\eta = 1/4$. However, the correlation length exponent $\nu$ varies continuously along the phase boundary. This variation is tightly constrained, in that $\nu$ obeys the Ashkin-Teller relation which connects it to the value of the anomalous exponent $\eta_2$ of the subdominant two-fold symmetry breaking field obtained by squaring the local four-fold symmetry breaking order parameter. This relation reads: $\eta_2(\nu) = 1-1/2\nu$.

In order to test this scenario, we need to identify this subdominant ordering field. From the symmetry considerations summarized above, this is straightforward to do: We see that the order parameter $\hat{\gamma}_{\rm P}$ has the same symmetry as the required two-fold symmetry breaking order parameter field, and therefore plays the role of this subdominant order parameter. With this motivation, we measure the anomalous exponents $\eta$ and $\eta_2$ associated with the critical behavior of the static correlator of $\hat{\psi}_{\rm B}$ and $\hat{\gamma}_{\rm P}$ respectively in addition to measuring the correlation length exponent $\nu$ that governs this critical behavior.

The results of such a study are shown in Fig.~\ref{fig:N3_AT_phi2,Qpipi} and Fig.~\ref{fig:N12_AT_phi2,Qpipi}.
In our analysis, we {\em use} the fact that $\eta =1/4$ to identify the critical temperature. With this identification made, we estimate $\nu$ by adjusting it to achieve the best scaling collapse in the vicinit of this critical point. With this in hand, we test for the validity of the Ashkin-Teller relation by measuring the anomalous exponent $\eta_2$ at this critical point. From the results displayed in these two figures, we see that $\nu$ indeed varies with parameters $N$ and $Q/J$, as does $\eta_2$. However, the Ashkin-Teller relation continues to hold.

\section{Discussion}
\label{sec:Discussion}
From a theoretical point of view, perhaps the most intriguing result of our study of these SU(N) analogs of $S=1$ antiferromagnets on the square lattice is the intermediate $(\pi,\pi)$-VBS ordered state identified here. What makes this state puzzling is the absence of any identifiable order in the bond or plaquette energies at wavevectors $(\pi, 0)$ and $(0, \pi)$. Indeed, we have been quite unable to come up with a simple caricature for this state in terms of a variational wavefunction that builds in this pattern of symmetry breaking. Identifying such a simple variational wavefunction is perhaps the most interesting question to emerge from our study. In particular, the absence of bond or plaquette correlations at wavevectors $(\pi, 0)$ and $(0, \pi)$ suggests the need to build into the wavefunction local resonances that destroy such ordering but preserve a checkerboard pattern of plaquette energies, corresponding to order at wavevector $(\pi,\pi)$.

\section{Acknowledgements}
  S.K. and N.D. were supported by graduate and postdoctoral fellowships of the Tata Institute of Fundamental Research (TIFR).
K.D. was supported at the
TIFR by DAE, India and in part by a J. C. Bose Fellowship
(No. JCB/2020/000047) of SERB, DST India, and by the
Infosys-Chandrasekharan Random Geometry Center (TIFR). The computational results presented here relied crucially on the generous allocation of computer resources of the Department of Theoretical Physics, TIFR, funded by DAE (India).
\bibliography{plqphase.bib}

\end{document}